# Decomposition analysis to identify intervention targets for reducing disparities

John W. Jackson and Tyler J. VanderWeele[1]

**ABSTRACT**

There has been considerable interest in using decomposition methods in epidemiology (mediation analysis) and economics (Oaxaca-Blinder decomposition) to understand how health disparities arise and how they might change upon intervention. It has not been clear when estimates from the Oaxaca-Blinder decomposition can be interpreted causally because its implementation does not explicitly address potential confounding of target variables. While mediation analysis does explicitly adjust for confounders of target variables, it does so in a way that entails equalizing confounders across racial groups, which may not reflect the intended intervention. Revisiting prior analyses in the National Longitudinal Survey of Youth on disparities in wages, unemployment, incarceration, and overall health with test scores, taken as a proxy for educational attainment, as a target intervention, we propose and demonstrate a novel decomposition that controls for confounders of test scores (measures of childhood SES) while leaving their association with race intact. We compare this decomposition with others that use standardization (to equalize childhood SES alone), mediation analysis (to equalize test scores within levels of childhood SES), and one that equalizes both childhood SES and test scores. We also show how these decompositions, including our novel proposals, are equivalent to causal implementations of the Oaxaca-Blinder decomposition.

---

John Jackson is Assistant Professor in the Departments of Epidemiology and Mental Health at the Johns Hopkins Bloomberg School of Public Health and Tyler VanderWeele is Professor in the Departments of Epidemiology and Biostatistics at the Harvard T.H. Chan School of Public Health. Correspondence to john.jackson@jhu.edu

**INTRODUCTION**

Health disparities are differences in health between socially advantaged vs. disadvantaged groups that are considered unnecessary and unjust.[1] Although national and local efforts have sought to reduce and eliminate racial/ethnic disparities in health over the past few decades, they have in many cases persisted.[2,3] Reducing racial/ethnic disparities requires that we understand how they arise and develop interventions to target the mechanisms that perpetuate them.[4]

Socioeconomic status (SES) in early life is often considered a primary driver of racial/ethnic disparities in adulthood.[5,6] Despite this, it is not always accounted for in studies of racial/ethnic disparities. Analyses by Fryer[7] in the National Longitudinal Survey of Youth, patterned after analyses by Johnson and Neal,[8] suggest that racial/ethnic disparities in wages, unemployment, incarceration, and health might be largely driven by disparities in education. After controlling for test score percentiles from the Armed Forces Qualifying Test, a measure of pre-market skills, black-white differences in log-wages decreased by 72% and the disparity in self-reported health vanished. Though provocative, these results were not adjusted for measures of familial SES in childhood and might be confounded. The disparity reductions might reflect the effect of equalizing childhood SES rather than test scores. This would potentially limit their value for developing evidence-based policy to reduce disparities in wages and health by targeting disparities in education.

More broadly, one might consider how disparities in wages and other outcomes would change by eliminating disparities in childhood SES vs. disparities in test scores. The potential outcomes framework can be used to estimate how disparities might change by intervening on either target.[9]

One might use standardization to ask how disparities might change upon equalizing the childhood SES distribution across race, or use mediation analysis to ask how disparities might change upon equalizing the test score distribution across race among those with the same childhood SES.[10] One might even consider a joint intervention to equalize both childhood SES and test scores. Unfortunately, as we will show, neither standardization nor mediation analysis answers the question that appeared to motivate the earlier analyses by Neal and Johnson or Fryer[7,8]: how might disparities in adult outcomes change if we removed disparities in educational test scores but not childhood SES? If disparities in adult outcomes could be considerably reduced through education, this could lead to encouraging policy considerations.

In a re-analysis of the Fryer data, we outline and demonstrate a novel decomposition method that estimates how well removing disparities in test scores, but not childhood SES, reduces disparities in adulthood outcomes. We compare these results to estimates under interventions to equalize childhood SES alone (standardization), to equalize test scores among children of the same SES (mediation analysis), and to equalize childhood SES and test scores together. Unlike these other approaches, our method appropriately adjusts for the confounding effects of childhood SES on test scores while leaving its association with race intact. Conceptually, it maps to a randomized trial of an intervention where the estimand is the association between race and the outcome.[11] In the main text, we implement the method using formulae derived under linear models for the outcome. In the Appendix, we provide non-parametric formulae as well as those for non-linear models. There, we show how standardization, mediation analysis, and our decomposition can be expressed as Oaxaca-Blinder decompositions[12,13] under certain conditions.

**EXAMPLE DATA AND STATISTICAL ANALYSES**

Our motivating example revisits analyses by Fryer.[7] Replicating those data, we extracted baseline and outcome data from the National Longitudinal Survey of Youth on black and white men who were ages 14-22 years at baseline in the 1979 cohort[14] (NLSY79), and ages 12-16 at baseline in the 1997 cohort[15] (NLSY97) in the United States. We used data from baseline surveys to define indicators of ascertained gender and race (1=black, 0=white), ethnicity (1=Hispanic, 0=non-Hispanic), and mixed race (1=mixed race, 0=single race; NLSY97 only). Hourly wage in 2006 US dollars was calculated as a weighted average across all current jobs in 2006 or 2007 (with proportion of total hours/week per job as weights), excluding possibly implausible wages below $1 or above $115 per hour, and log-transformed. Unemployment was coded as a binary variable from current employment status in 2006, with individuals not in the labor market coded as missing. Incarceration was coded as a binary variable indicating self-reported residence in jail for any follow-up survey through 2006 or having been sentenced to a correctional institution before baseline. Self-reported health (only measured in NLSY79) was recoded from the physical component score from the 12-item Short Form Health Survey in 2006 and converted to a z-score. These represent outcomes in 2006 or 2007 at ages 42 through 44 for the NLSY79 cohort, and ages 22 through 27 for the NLSY97 cohort. Test score percentiles were obtained for the NLSY79 cohort from the Armed Forces Qualification Test—the sum of the arithmetic reasoning score, the mathematics knowledge score, and two times the verbal composite score—which was administered as part of the Armed Services Vocational Aptitude Battery as reported in the 1981 survey year. For NLSY97, the AFQT percentiles obtained were based on a similarly constructed (but unofficial) score from the 1999 survey year. Scores were standardized by age within the NLSY79 cohort, and also within the NLSY97 cohort, as described elsewhere.[7] Total years of education before 2006 or 2007 was also extracted, bottom-coded at 8 years and top-coded at 16 years. Measures of childhood SES in the NLSY79 cohort included maternal educational attainment (highest grade completed), household income, and poverty status as assessed in 1979. For the NLSY97 cohort we used the same measures of childhood SES (replacing poverty status with household net worth) which were assessed in 1997 or 1998. For each cohort, we extracted the first principal component from the measures of childhood SES. Missing indicators (1=missing, 0=otherwise) were constructed for test scores, total years of education, and all measures of childhood SES. See Fryer[7] for more details. The characteristics of the NLSY79 are described in Table 1, and the NLSY97 described in Appendix Table 1.

In the main text, we present formulae under models that rely on a single measure of childhood SES (e.g. the first principal component) because the formulae are more intuitive. Nonetheless, the disparity estimates reported in Tables 2 and 3 were obtained under models that relied on three separate measures of childhood SES. The formulae used are given in the eAppendix. Replicating Fryer,[7] all models included mutually exclusive dummy variables for Hispanic ethnicity and mixed race (for NLSY97 only) as well as missing indicators for education and childhood SES variables. The non-parametric bootstrap with 1,000 replication samples was used to obtain standard errors. The proportion of the disparity reduced was estimated on the additive scale (see eAppendix). Note that for the NLSY79 cohort we do not provide results for unemployment as there were insufficient cases for reliable estimates under the bootstrap.

**THE STRUCTURE OF EXTANT DISPARITIES IN ADULTHOOD**

Figure 1A portrays the structure of disparities as relationships between race R (1=black, 0=white), childhood SES X, test scores M, an outcome Y, covariates gender and age C, and historical processes H such as slavery and Jim Crow that are responsible for black-white differences in socioeconomic status and residence near conception.[16] The diagram could be detailed further by breaking the race node R into features that investigators might consider under their study, including skin color, its perception by others, cultural context, genetic background.[10] We retain a general race node R because our results apply to any definition of race that investigators use to operationalize the construct of race. The formal results in the Appendix are, however, given without reference to any particular causal diagram so the diagram is presented primarily for intuition.

In Figure 1A, the racial disparity in outcome Y arises in several ways. The disparity arises through backdoor paths involving history H: the effects of Jim Crow have been that blacks are more likely to be born into families with low socioeconomic status who live in neighborhoods with lower quality schools[16] (a non-mediating path). Forward paths emanating from the race node could represent effects of discrimination: blacks are more likely to be placed into less rigorous curriculum tracks in early education and the effects of this accumulate e.g. mathematics course choice in high school[17,18] (a mediating path). The direct path is comprised of all forward pathways that do not operate through variables considered here.

**DISPARITY REDUCTIONS UNDER ALTERNATIVE INTERVENTION STRATEGIES**

We now describe results from decompositions that estimate how well certain interventions might reduce racial disparities in adulthood (wages, unemployment, incarceration, and health) by equalizing childhood SES and/or test scores across race. Each intervention is equivalent to deactivating certain paths linking race to adult outcomes. To identify the disparity reductions with observational data, we assume: (A1) the effect of childhood SES on an outcome is unconfounded given race and covariates (such as gender and age); (A2) the effect of test scores on an outcome is unconfounded given race, childhood SES, and covariates.

We report estimates for the initial disparity, the residual after each intervention, and the corresponding reduction along with their standard errors in Table 2 (regarding interventions for test scores) and Table 3 (regarding interventions for total years of education) for the NLSY79 cohort. Our narrative focuses on log-wages to introduce the method and later summarizes results for incarceration and self-reported health. Results for the NLSY97 cohort are provided in the Appendix.

*Proposition 1: Intervene to Equalize Childhood SES across Race*

The first proposal is to randomly assign childhood SES among blacks such that they follow the distribution among whites of the same gender and age. We posit that childhood SES reflects conditions near the time of conception and thus link race with adult outcomes through a backdoor path e.g. R ←H→X→Y in Figure 1A. VanderWeele and Robinson[10] provided analytic formulae for the residual disparity under equalizing a non-mediating variable such as childhood SES. These formulae require that assumption A1 holds. Provided this, we can estimate the residual disparity by fitting linear models that condition on race R and covariates gender and age C (1) and additionally childhood SES X (2):

$$E[Y|r,c] = \phi_0 + \phi_1 r + \phi_4'c \quad (1)$$

$$E[Y|r,x,c] = \gamma_0 + \gamma_1 r + \gamma_2 x + \gamma_4'c \quad (2)$$

where $\phi_1$ represents the overall race disparity in log wages, $\gamma_1$ represents the race disparity given childhood SES, both conditional on covariates. Under an intervention to equalize the distribution of childhood SES across race, the residual disparity would equal $\gamma_1$, and the disparity reduction would be $\phi_1-\gamma_1$. Thus, the initial disparity $\phi_1$ which equals -0.41 (0.04) would under proposition 1 decrease to $\gamma_1$ which equals -0.30 (0.05), a 26% reduction. Figure 1B shows that proposition 1 corresponds to deactivating the backdoor path between race and childhood SES. Although the results here are illustrated using linear models, all of the approaches can be implemented in more general settings and non-parametric results are given in the Appendix.

*Proposition 2: Intervene to Equalize Test Scores within Levels of Childhood SES across Race*
The second proposal is to randomly assign educational attainment (reflected in test scores) among blacks such that they follow the same distribution among whites of the same gender, age, and childhood SES (note that the economics literature refers to AFQT test scores as measures of pre-market skills). This intervention attempts to remove disparities in test scores that cannot be attributed to disparities in childhood SES (a mediating path); it is not concerned with eliminating disparities that operate through childhood SES (a backdoor path). VanderWeele and Robinson[10] provided analytic formulae for the residual disparity under conditionally equalizing a mediating variable such as test scores. These formulae require that assumption A2 holds. If this is so, we could estimate the residual disparity by fitting linear models that condition on race R, childhood SES X, covariates gender and age C (2) and additionally test scores M (3):

$$E[Y|r,x,c] = \gamma_0 + \gamma_1 r + \gamma_2 x + \gamma_4'c \tag{2}$$

$$E[Y|r,x,m,c] = \theta_0 + \theta_1 r + \theta_2 x + \theta_3 m + \theta_4'c \tag{3}$$

where $\gamma_1$ represents the race disparity given childhood SES, and $\theta_1$ represents the disparity upon further stratifying on test scores, both of which condition on covariates. Under an intervention to take children with the same childhood SES and equalize their test score distribution across race, the residual disparity would equal $\theta_1$, and the disparity reduction would be $\gamma_1 - \theta_1$. Thus, the initial disparity $\gamma_1$ which equals -0.30 (0.05) would decrease to $\theta_1$ which equals -0.11 (0.05). Because this 65% reduction pertains to children who share the same childhood SES, much of the marginal disparity, without conditioning on childhood SES, between blacks and whites would remain. Figure 1C shows that the intervention eliminates the mediated path involving test scores but leaves backdoor paths involving childhood SES intact.

*Proposition 3: Intervene to Equalize both Test Scores and Childhood SES across Race*

The third proposal is to randomly assign childhood SES and educational attainment (reflected in test scores) among blacks such that they follow the same distribution among whites. This intervention targets mediated paths and back-door paths by which disparities in log-wages arise. In the Appendix we extend the results of VanderWeele and Robinson[10] to provide formulae for the residual disparity under jointly equalizing a non-mediating variable such as childhood SES and also a mediating variable such as test scores. These formulae require that both assumptions A1 and A2 hold. Provided this is so, we could estimate the residual disparity by fitting linear models that condition on race R, covariates gender and age C (1) and additionally childhood SES X and test scores M (3):

$$E[Y|r,c] = \phi_0 + \phi_1 r + \phi_4'c \qquad (1)$$

$$E[Y|r,x,m,c] = \theta_0 + \theta_1 r + \theta_2 x + \theta_3 m + \theta_4'c \qquad (3)$$

where $\phi_1$ represents the overall race disparity in log wages, and $\theta_1$ represents the disparity upon further stratifying on childhood SES and test scores, both of which condition on covariates. Under an intervention to equalize both the childhood SES and test score distributions across race, the residual disparity would equal $\theta_1$, and the disparity reduction would be $\phi_1-\theta_1$. Thus, the initial disparity $\gamma_1$ which equals -0.41 (0.04) would decrease to $\theta_1$ which equals -0.11 (0.05), a 74% reduction. Figure 1D shows that the intervention eliminates the mediated path involving test scores as well as the backdoor paths involving childhood SES.

*Proposition 4: Intervene to Equalize Test Scores across Race*
The previous propositions focused on targeting backdoor vs. mediated paths that generate the disparity in log-wages. But there is a conceptual issue with proposition 2. Identifying the effect of eliminating a mediated path (through test scores) requires adjustment of confounders (of test scores e.g. childhood SES, gender and age). This may be problematic because achieving disparity reductions for children *of the same childhood SES* may constrain black children with low SES to a test score distribution that is already suboptimal. Test score disparities that arise through disparities in childhood SES would persist. These issues arise regardless of how one adjusts for childhood SES, because methods for indirect effects will always statistically equalize the distribution of a non-mediating variable across race through stratification or standardization.

An alternative is to shift our focus away from eliminating mediated paths involving test scores and towards eliminating disparities in test scores entirely, regardless of whether they arise through mediated or backdoor paths. This describes the fourth proposition, which is to randomly assign educational attainment (reflected in test scores) among blacks such that they follow the same marginal distribution as among whites. In the Appendix, we provide formulae for the residual disparity under equalizing a variable such as test scores that lies along a mediating path and also a non-mediating path. These formulae adjust for confounding of test scores by childhood SES, but preserve the relationship between race and childhood SES. They require that assumption A2 holds. Provided this is so, we could estimate the residual disparity by fitting linear models that condition on race R, covariates gender and age C (1) and additionally upon childhood SES X (2) and finally test scores M (3):

$$E[Y|r,c] = \phi_0 + \phi_1 r + \phi_4'c \tag{1}$$

$$E[Y|r,x,c] = \gamma_0 + \gamma_1 r + \gamma_2 x + \gamma_4'c \tag{2}$$

$$E[Y|r,x,m,c] = \theta_0 + \theta_1 r + \theta_2 x + \theta_3 m + \theta_4'c \tag{3}$$

where $\phi_1$ represents the overall race disparity in log wages, $\gamma_1$ represents the disparity given childhood SES, $\theta_1$ represents the disparity given childhood SES and test scores, $\gamma_2$ represents the race-specific total effect of childhood SES on log-wages, and $\theta_2$ represents the race-specific direct effect of childhood SES on log-wages (with respect to test scores) which all condition on covariates. Under an intervention to equalize test scores alone across race, the residual disparity would equal $\theta_1+(\theta_2/\gamma_2)(\phi_1-\gamma_1)$, and the disparity reduction would be $(\gamma_1-\theta_1)+(1-\theta_2/\gamma_2)(\phi_1-\gamma_1)$. Thus, the initial disparity, which equals -0.41 (0.04), would decrease to -0.14 (0.10), a 66% reduction. Conceptually, the disparity reduction estimate begins with the decrease that occurs under equalizing test scores within levels of childhood SES ($\gamma_1-\theta_1$). To this amount, we add in the disparity reduction under equalizing childhood SES alone ($\phi_1-\gamma_1$) but only scaled by the proportion that is mediated by test scores ($1-\theta_2/\gamma_2$); this accounts for the extent to which an intervention on test scores would block the effect of childhood SES on the outcome. If test scores does not mediate the effect of childhood SES, such that $\theta_2=\gamma_2$, none of the disparity reduction under equalizing childhood SES alone is added and the expression simplifies to the disparity reduction under proposition 2 i.e. $\gamma_1-\theta_1$. If test scores completely mediates the effect of childhood SES, such that $\theta_2=0$, then all of the disparity reduction under equalizing childhood SES alone is added, and the expression simplifies to the reduction under proposition 3 i.e. $\phi_1-\theta_1$. In Figure 1E we see that the intervention removes a mediating path and a backdoor path that involve test scores, and that removing the backdoor path involves equalizing test scores across childhood SES.

**A RE-ANALYSIS OF FRYER**

In the striking results of Fryer, black-white disparities in log-wages, unemployment, and health were substantially reduced upon controlling for test scores. We reconstructed the Fryer analyses which did not adjust for childhood SES (by omitting the term for X in model (3)) and compared these results from those obtained under the four propositions described above that do account for childhood SES. In Table 2, the NLSY79 disparity in log-wages was -0.41 (0.04), a quarter (26%) of which would be removed under equalizing childhood SES alone (proposition 1). Equalizing childhood SES and test scores together (proposition 3) would remove nearly three-quarters (74%) of the disparity. Interventions to equalize test scores alone would remove two-thirds (66%) of the disparity. For the disparity in incarceration, 3.54 (1.17), equalizing childhood SES would remove just over two-fifths (45%) of the disparity, equalizing both childhood SES and test scores would reduce four-fifths (81%) of the disparity, and equalizing test scores alone would remove two-thirds of the disparity (65%). For the disparity in self-reported health -0.14 (0.05), equalizing childhood SES alone would remove all of the disparity (taking the confidence intervals into account) and this would be true for interventions to equalize childhood SES and test scores together and also interventions to equalize test scores alone.

For each outcome, equalizing both childhood SES and test scores would be most effective. Nonetheless, equalizing test scores alone would be nearly as effective, and sometimes more effective than equalizing socioeconomic status in early life. The findings in Table 3 show, however, that the nature of the hypothetical educational intervention is critical. Repeating the analyses to equalize total years of education rather than test scores under proposition 4 showed much smaller reductions of 27% in log-wages, 13% in incarceration, and 16% in health. This may be because the racial gaps were wider for test scores than for total years of education, and this is worth further study. Overall, our results under proposition 4 were qualitatively similar to those of Fryer. Skills obtained through educational attainment (as measured by test scores), rather than total years of education, may be a more attractive target for substantially reducing disparities.

Our analyses are limited in several respects. Although we used multiple measures to capture childhood SES, our results are still subject to residual confounding from measurement error and missing data, and unmeasured confounding by dimensions of SES there were not included (e.g. parental occupation).[19,20] Our ability to account for SES is also limited by the fact that increases in SES do not necessarily translate into the same gains for blacks and whites.[21] Furthermore, our analyses could employ sampling weights to account for the NLSY design and this could be explored in future empirical work. Despite these caveats, the results' magnitude underscore how addressing education-based disparities is likely a key component in eliminating disparities in economic opportunity, justice, and health.

**DISCUSSION**

We have presented a way to decompose an extant disparity in adult outcomes (e.g. log-wages) into a reduction and a residual portion upon equalizing disparities in a target variable that lies on a mediating path (e.g. test scores), even when that target is confounded by a variable that lies on a non-mediating path (e.g. childhood SES). This approach appropriately controls for confounding by a variable like childhood SES in a way that preserves its relationship with race. This feature overcomes a conceptual constraint of current methods for indirect effects which can only estimate how well interventions reduce disparities after statistically equalizing confounding variables such as childhood SES across race. Substantively, interventions that intend to align educational outcomes within groups whose educational outcomes are already on average suboptimal (e.g. those with low SES) may not be effective.

Also of interest, substantively, was our finding that equalizing test scores alone would be nearly as effective, and sometimes more effective, than equalizing socioeconomic status in early life. This is an important finding because achieving equity in education across socioeconomic status across race and class (which is what proposition 4 implies) may be a more feasible intervention, in the short run at least, than eliminating racial differences in childhood socioeconomic conditions. Possible ways to accomplish this could include, among state and county-level initiatives to improve quality and cost-efficiency, de-linking public school funding from local property tax revenue, or increasing federal funding to schools with high concentrations of low-income students and high achievement gaps.[22]

Another important point with substantive implications is that estimating disparity reductions while equalizing confounding variables could lead to misinterpretations. This can be seen perhaps most clearly when examining years of education rather than test scores. In Table 3, a mediation analysis that equalized total years of education within levels of childhood SES (proposition 2) gave roughly the same numeric values for the residual disparity as an analysis that equalized childhood SES and test scores jointly (proposition 3), and these were much smaller than the residual disparity under equalizing test scores marginally (proposition 4). But the residual disparity under proposition 2 ignores disparities that remain through race's association with childhood SES. Without careful interpretation of results from the mediation analysis, one might over-interpret the importance of total years of education for reducing disparities. Our decomposition method does not suffer from these limitations.

There has been considerable debate in the statistics, social science, and epidemiology literature as to whether socially defined characteristics such as race can be given causal attribution and whether their effects can be identified from observational data.[23–30] Our contribution is not meant to advance this debate. Our decomposition does not focus on the causal status of race or attempt to identify its effect. Rather it focuses causal inference on potentially manipulable targets and their ability to reduce the association between race and an outcome.[9] Mediation analysis methods—even when reframed to adopt this viewpoint[10,31]—requires epidemiologists to first equalize confounding variables across race and only then consider what targets can reduce that adjusted disparity.[32] Our approach sidesteps this restriction and opens a broader range of inquiry for reducing disparities.

Our contribution has implications for other approaches used to understand disparities. In economics, sources of disparities are often identified using the Oaxaca-Blinder decomposition. This method disaggregates the disparity into a portion due to statistical variation in the covariates—the explained portion—and an unexplained portion that is usually attributed to discrimination.[12,13] The Oaxaca-Blinder decomposition is increasingly appearing the public health literature,[33,34] sometimes with causal interpretation e.g. with the explained portion described as the disparity reduction under an intervention to equalize risk factors (targets). These interpretations are highly questionable when they do not explicitly account for how those targets may be confounded.[35] When all confounders of targets are adjusted for, and moreover that confounders of targets in later life are not affected by race or targets in early life (i.e. no time-dependent confounding), each of propositions 1 through 4 can be accomplished as an Oaxaca-Blinder decomposition (see Appendix). When there is time-dependent confounding, the Oaxaca-Blinder Decomposition methods, and also the parametric formulae in the main text would be vulnerable to selection-bias[10,36] and should not be used. In the eAppendix, we present non-parametric formulae that can be used to implement propositions 2, 3 and 4 in the presence of a time-dependent confounder, effectively also generalizing Oaxaca-Blinder decomposition methods to this setting as well.

The results presented above require that the models be correctly specified; in the main text these formulas do not, for example, account for possible interaction between race, childhood and test scores. However, the general non-parametric results given in the eAppendix can be used to derive estimators for Propositions 2, 3 and 4 that allow not only for time-dependent confounding but also for interactions and less sensitivity to modeling assumptions, as has been done elsewhere.[37] The non-parametric formulae still require that confounders of targets be measured and adjusted for. It will be important in future research to develop intuitive sensitivity analyses that can quantify potential bias when some confounders are unmeasured.[38] Future research could also expand this method to consider disparity reductions along multiple axes of disadvantage beyond race i.e. questions framed with an intersectional focus.[39,40]

We have introduced a new perspective on how to use the potential outcomes framework to identify targets that appear attractive for reducing disparities. We hope these methods enable epidemiologists to help advance research priorities, policy initiatives, and intervention design to eliminate health disparities.

multiple robustness, and sensitivity analysis. *Ann Stat*. 2012;40(3):1816-1845.

38. Ding P, Vanderweele TJ. Sharp sensitivity bounds for mediation under unmeasured mediator-outcome confounding. *Biometrika*. 2016;103(2):483-490.

39. Jackson JW, Williams DR, VanderWeele TJ. Disparities at the intersection of marginalized groups. *Soc Psychiatry Psychiatr Epidemiol*. 2016;51(10):1349-1359.

40. Jackson JW. Explaining Intersectionality through description, counterfactual thinking, and mediation analysis. *Soc Psychiatry Psychiatr Epidemiol*. 2017 (under review).
20

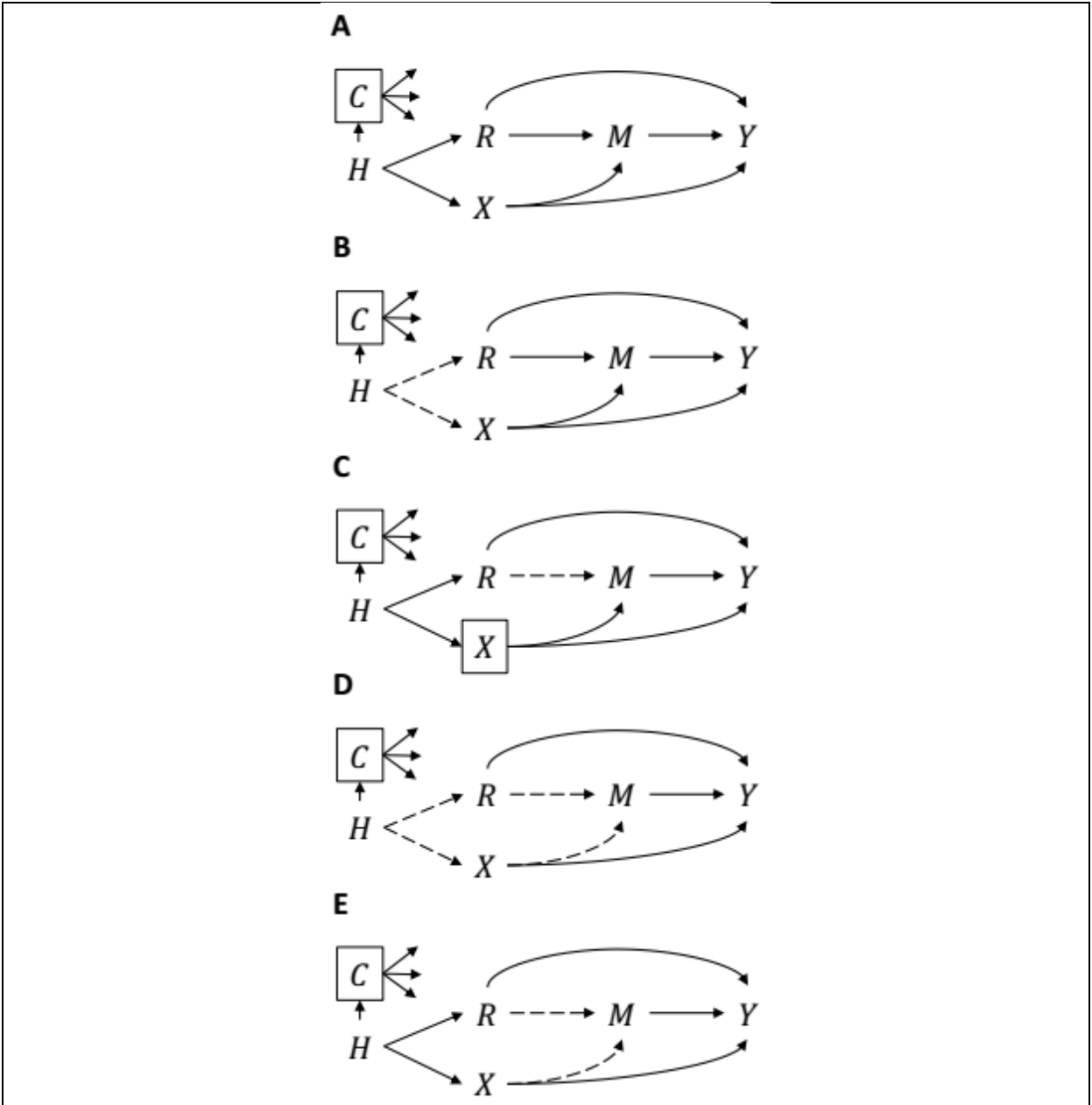

Figure 1. Diagram representing relationships between race R, an outcome Y, measures of characteristics in early life e.g. childhood SES X, measures of characteristics later in life e.g. test scores M, history H, and covariates gender and age C in the population (a) and under alternative interventions (b)-(e), wherein solid arrows represent relationships that are preserved and dashed arrows represent relationships that are abolished. In (b) an intervention under proposition 1 to equalize childhood SES across race deactivates the backdoor paths R ←H→X→M→Y and R ←H→X→ Y. In (c) an intervention under proposition 2 to equalize test scores within levels of childhood SES deactivates only the mediated path R→M→Y. In (d) an intervention to equalize both childhood SES and test scores across race deactivates the mediated path R→M→Y and also the backdoor paths R ←H→X→M→Y and R ←H→X→ Y. In (e) an intervention to equalize test scores marginally across race deactivates the mediated path R→M→Y and the backdoor path R ←H→X→M→Y, but leaves the backdoor path R←H→X→Y intact.

| Table 1. Characteristics of males in the 1979 National Survey of American Youth Analytic Cohort, mean (standard error) | | |
|---|---|---|
| | White (n=1,010) | Black (n=597) |
| Age | 43.1 (0.8) | 43.1 (0.8) |
| | | |
| Adulthood outcomes | | |
|   Wage (dollars / hr) | 26.1 (17.4) | 17.4 (12.2) |
|   Unemployed (%) | 3.6 (18.6) | 8.0 (27.2) |
|   Incarceration, ever (%) | 7.4 (26.2) | 22.1 (41.5) |
|   Standardized physical component score | 0.15 (0.8) | 0.3 (1.0) |
| | | |
| Measures of educational attainment | | |
|   Armed Forces Qualifying Test (AFQT) | 0.45 (1.0) | -0.58 (0.8) |
|   Total years education | 13.3 (2.1) | 12.6 (1.8) |
| | | |
| Measures of childhood SES | | |
|   Mother's highest grade level | 11.9 (2.4) | 10.9 (2.5) |
|   Poverty Status in childhood (%) | 9.6 (29.5) | 48.7 (50.0) |
|   Household Income in childhood | $21,466 ($12,854) | $10,835 ($7799) |
| | | |
| Proportion missing (%) | | |
|   Missing AFQT | 3.9 (19.3) | 2.2 (14.6) |
|   Missing total years of education | 25.6 (43.7) | 23.3 (42.3) |
|   Missing mother's highest grade level | 5.3 (22.5) | 10.2 (30.3) |
|   Missing poverty status in childhood | 9.4 (39.1) | 4.4 (20.4) |
|   Missing household income in childhood | 18.8 (39.1) | 17.3 (37.8) |

| | Proposition 1 | Proposition 2 | Proposition 3 | Proposition 4 | Re-analysis of Fryer |
|---|---|---|---|---|---|
| | Intervene to equalize the distribution of childhood SES measures across race but not AFQT scores | Intervene to equalize the distribution of AFQT scores across race within levels of childhood SES | Intervene to equalize the distribution of AFQT scores and childhood SES measures across race | Intervene to equalize the distribution of AFQT scores across race but not childhood SES measures | Statistically equalize the distribution of AFQT scores across race without control for childhood SES |
| **Log wages** | | | | | |
| Initial disparity | -0.41 (0.04) | -0.30 (0.05) | -0.41 (0.04) | -0.41 (0.04) | -0.41 (0.04) |
| Residual disparity | -0.30 (0.05) | -0.11 (0.05) | -0.11 (0.05) | -0.14 (0.10) | -0.13 (0.05) |
| % reduction | 26 | 65 | 74 | 66 | 69 |
| **Incarceration** | | | | | |
| Initial disparity | 3.54 (1.17) | 2.39 (1.21) | 3.54 (1.17) | 3.54 (1.17) | 3.54 (1.17) |
| Residual disparity | 2.39 (1.21) | 1.49 (1.21) | 1.49 (1.21) | 1.86 (1.25) | 1.76 (1.19) |
| % reduction | 45 | 65 | 81 | 65 | 70 |
| **Health** | | | | | |
| Initial disparity | -0.14 (0.05) | -0.04 (0.06) | -0.14 (0.05) | -0.14 (0.05) | -0.14 (0.05) |
| Residual disparity | -0.04 (0.06) | 0.05 (0.07) | 0.05 (0.07) | -0.02 (0.07) | 0.02 (0.06) |
| % reduction | 75 | 251 | 137 | 85 | 112 |

Table 2. Estimates of residual disparities and disparity reductions in adult outcomes under hypothetical intervention strategies on childhood SES measures and/or Armed Forces Qualifying Test scores in the 1979 NLSY Cohort[1]

[1]The analytic sample size was 1154 for wages, 1988 for incarceration, and 1587 for health. All models included a mutually exclusive dummy variable for Hispanic ethnicity.

Table 3. Estimates of residual disparities and disparity reductions in adult outcomes under hypothetical intervention strategies on childhood SES measures and/or total years of education in the 1979 NLSY Cohort[1]

| | Proposition 1 | Proposition 2 | Proposition 3 | Proposition 4 | Re-analysis of Fryer |
|---|---|---|---|---|---|
| | Intervene to equalize the distribution of childhood SES measures across race but not total years of education | Intervene to equalize the distribution of total years of education across race within levels of childhood SES | Intervene to equalize the distribution of total years of education and childhood SES measures across race | Intervene to equalize the distribution of total years of education across race but not childhood SES measures | Statistically equalize the distribution of total years of education across race without control for childhood SES |
| Log wages | | | | | |
|   Initial disparity | -0.41 (0.04) | -0.30 (0.05) | -0.41 (0.04) | -0.41 (0.04) | -0.41 (0.04) |
|   Residual disparity | -0.30 (0.05) | -0.26 (0.05) | -0.26 (0.05) | -0.30 (0.04) | -0.30 (0.04) |
|   % reduction | 26 | 15 | 37 | 27 | 28 |
| | | | | | |
| Incarceration | | | | | |
|   Initial disparity | 3.53 (1.17) | 2.39 (1.21) | 3.53 (1.17) | 3.53 (1.17) | 3.53 (1.17) |
|   Residual disparity | 2.39 (1.21) | 2.36 (1.21) | 2.36 (1.21) | 3.22 (1.45) | 3.06 (1.18) |
|   % reduction | 45 | 2 | 46 | 13 | 19 |
| | | | | | |
| Health | | | | | |
|   Initial disparity | -0.14 (0.05) | -0.03 (0.06) | -0.14 (0.05) | -0.14 (0.05) | -0.14 (0.05) |
|   Residual disparity | -0.03 (0.06) | -0.03 (0.06) | -0.03 (0.06) | -0.11 (0.22) | -0.11 (0.05) |
|   % reduction | 75 | 12 | 78 | 16 | 24 |

[1]The analytic sample size was 1154 for wages, 1988 for incarceration, and 1587 for health. All models included a mutually exclusive dummy variable for Hispanic ethnicity.

# PRINT APPENDIX

## Introduction and Notation

Consider a comparison of two race/ethnicity groups and let R denote a binary variable indicating race. Let X be a set of characteristics at birth or early childhood that are potentially manipulable (e.g. early SES measures), let M be one or more characteristics later in life or in adulthood that are potentially manipulable (e.g. educational attainment or adult SES), let Y be some outcome of interest and let C be some other set of covariates at birth (e.g. gender, year-of-birth/age). The overall disparity measure within strata of covariates C (gender and age) would then be $E[Y|R=1,c]-E[Y|R=0,c]$. Unless noted otherwise, we will consider X to be a single measure of characteristics at birth.

Let $Y(x)$ be the value of the outcome that would have been observed for an individual had X been set to x. Likewise let $Y(m)$ be the value of the outcome that would have been observed for an individual had M been set to m. Finally let $Y(x,m)$ be the value of the outcome that would have been observed for an individual had X been set to x.

Unless otherwise noted we will assume:
A1: The effect of X on the outcome Y is unconfounded given (R,C)
A2: The effect of M on the outcome Y is unconfounded given (R,C,X)

## Nonparametric results

Here we give non-parametric results for each of the various decompositions in the absence of time-dependent confounding. Estimates that are obtained from linear or logistic models from each of the decompositions are summarized in Tables 4 and 5. Non-parametric results in the presence of time-dependent confounding can be found in the eAppendix.

Proposition 1 (VanderWeele and Robinson, 2014). The disparity that would remain if the childhood distribution of X for black persons (R=1) with covariates C=c were set equal to its distribution for white persons (R=0) with C=c would be:
$\mu_x - E[Y|R=0,c]$
and the amount the disparity is reduced would be:
$E[Y|R=1,c] - \mu_x$
where $\mu_x = \Sigma_x E[Y|R=1,x,c] P(x|R=0,c)$.

Proposition 2 (VanderWeele and Robinson, 2014). The disparity that would remain if the distribution of M for black persons (R=1) with covariates C=c and X=x were set equal to its distribution for white persons (R=0) with C=c and X=x would be:
$\mu_{m|x} - E[Y|R=0,x,c]$
and the amount the disparity is reduced would be:
$E[Y|R=1,x,c] - \mu_{m|x}$
where $\mu_{m|x} = \Sigma_m E[Y|R=1,x,m,c] P(m|R=0,x,c)$.

Proposition 3. The disparity that would remain if the distribution of (X,M) for black persons (R=1) with covariates C=c were set equal to its distribution for white persons (R=0) with C=c would be:
$\mu_{xm} - E[Y|R=0,c]$
and the amount the disparity is reduced would be:

E[Y|R=1,c]- $\mu_{xm}$
where $\mu_{xm} = \Sigma_{x,m}$ E[Y|R=1,x,m,c]P(m|R=0,x,c)P(x|R=0,c).

Proposition 4. The disparity that would remain if the distribution of M for black persons (R=1) with covariates C=c were set equal to its distribution for white persons (R=0) with C=c would be:
$\mu_m$ -E[Y|R=0,c]
and the amount the disparity is reduced would be:
E[Y|R=1,c]- $\mu_m$
where $\mu_m = \Sigma_{x,m}$ E[Y|R=1,x,m,c]P(m|R=0,c)P(x|R=1,c).

### Results expressed as Oaxaca-Blinder decompositions

The Oaxaca-Blinder decomposition[12,13] is often used in labor economics to understand how much differences in group-characteristics explain disparities (or differences, more generally) in outcomes across groups e.g. disparities in log-wages across women vs. men, blacks vs. whites, union-members vs. non-members, etc. It partitions the total wage-difference into a portion due to differences in the distribution of potentially explanatory variables (termed the explained portion or composition effect), and a residual portion that cannot be explained by differences in these variables (termed the unexplained portion or structure effect). The unexplained portion is referred to as the structure effect because, as we will show, it captures the extent to which associations between the explanatory variables and the outcome vary across groups. Though the Oaxaca-Blinder decomposition was first introduced using linear models to decompose mean differences, more general forms have been introduced to decompose non-linear outcomes and even the entire distribution of the outcome. Here we outline the Oaxaca-Blinder decomposition under linear models for the mean and discuss the conditions under which the Propositions 1 through 4 can be viewed as a causal version of the Oaxaca-Blinder decomposition with respect to interventions to set the distributions of the explanatory variables. Previous literature has concerned causal inference with respect to interventions to set group membership e.g. race, gender, union-membership etc.[35] but the interpretation of such an intervention is more difficult when the variable to be intervened upon is race or gender. We consider an alternative causal interpretation below.

<u>A review of marginal Oaxaca-Blinder decompositions</u>

Let us consider an Oaxaca-Blinder decomposition to estimate the portion of the racial disparity in log-wages Y that is statistically explained vs. not explained by racial differences in variables $V_1...V_n$, where the comparison across race concerns blacks R=1 vs. whites R=0. A typical Oaxaca-Blinder decomposition would proceed by fitting two race-specific regressions for the outcome given the explanatory variables:

$E[Y|R = 1, v] = \beta_0^{R=1} + \beta_1^{R=1} v_1 + \beta_2^{R=1} v_2 + ... + \beta_n^{R=1} v_n$
$E[Y|R = 0, v] = \beta_0^{R=0} + \beta_1^{R=0} v_1 + \beta_2^{R=0} v_2 + ... + \beta_n^{R=0} v_n$

Along with these we would estimate the mean value of each explanatory variable $V_j$ among whites e.g. $E[V_1|R=0],...,E[V_n|R=0]$. Then, the typical Oaxaca-Blinder decomposition expresses the marginal racial disparity in mean log-wages as a function of the explanatory variables' means (among whites) and their race-specific regression parameters:

E[Y|R=1]- E[Y|R=0]
$= (\beta_0^{R=1} - \beta_0^{R=0}) + \sum_{j=1}^{n}(\beta_1^{R=1} - \beta_0^{R=0})E[V_j|R = 0] + \sum_{j=1}^{n} \beta_j^{R=1}\{E[V_j|R = 1] - E[V_j|R = 0]\}$

In what is called the aggregate decomposition,[35] the goal is to understand the extent to which the racial disparity is statistically explained by the fact that racial groups have different means for the explanatory variables. The term $\sum_{j=1}^{n}\beta_j^{R=1}\{E[V_j|R=1]-E[V_j|R=0]\}$ comprises the 'explained portion' or what is also called the 'composition effect.' It captures racial differences in the mean values of the explanatory variables. The sum of the terms $(\beta_0^{R=1}-\beta_0^{R=0})$ and $\sum_{j=1}^{n}(\beta_j^{R=1}-\beta_j^{R=0})E[V_j|R=0]$ comprises the 'unexplained portion' or what is also called the 'structure effect.' It captures the portion of the disparity that cannot be statistically explained by differences in the means of explanatory variables i.e. differences in mean log-wages at the reference levels of the explanatory variables and also racial differences in the associations between each explanatory variable and the mean of the outcome log-wages.

In what is called the detailed decomposition,[35] the term $\beta_j^{R=1}\{E[V_j|R=1]-E[V_j|R=0]\}$ is interpreted as the independent contribution of the explanatory variable $V_j$ to the 'explained portion' i.e. the portion of the disparity that is statistically attributable to the fact that the mean of $V_j$ differs across racial groups (independently of racial differences in the means of the other explanatory variables). The term $\sum_{j=1}^{n}(\beta_j^{R=1}-\beta_j^{R=0})E[V_j|R=0]$ is interpreted as the contribution of $V_j$ to the 'unexplained portion' i.e. the portion of the disparity that is statistically explained by differences in the race-specific associations between the explanatory variable $V_j$ and the mean of log-wages.

Defining conditional Oaxaca-Blinder decompositions

The typical Oaxaca-Blinder decomposition concerns the marginal racial disparity in log-wages E[Y|R=1]- E[Y|R=0], but one can extend it to decompose the racial disparity within levels of conditioning variables C i.e. E[Y|R=1,c]- E[Y|R=0,c]. These conditioning variables differ from explanatory variables $V_j$ in that they are used to define the population rather than to explain the disparity. To accomplish this, one first fits race-specific models for the mean of log-wages given the explanatory variables $V_j$ and also the conditioning variables C.

$$E[Y|R=1,v,c] = \beta_0^{R=1,c} + \beta_1^{R=1,c}v_1 + \beta_2^{R=1,c}v_2 + \ldots + \beta_n^{R=1,c}v_n + \beta_c^{R=1,c}c$$
$$E[Y|R=0,v,c] = \beta_0^{R=0,c} + \beta_1^{R=0,c}v_1 + \beta_2^{R=0,c}v_2 + \ldots + \beta_n^{R=0,c}v_n + \beta_c^{R=0,c}c$$

It can be shown that, the disparity within levels of C can be expressed as a function of the means of explanatory variables $V_j$ given C=c and also the regression parameters that also condition on C=c:

E[Y|R=1,c]- E[Y|R=0,c]
$= (\beta_0^{R=1,c} - \beta_0^{R=0,c}) + \sum_{j=1}^{n}(\beta_j^{R=1,c} - \beta_j^{R=0,c})E[V_j|R=0,c] + \sum(\beta_{c'}^{R=1,c} - \beta_{c'}^{R=0,c})c' + \sum_{j=1}^{n}\beta_j^{R=1,c}\{E[V_j|R=1,c] - E[V_j|R=0,c]\}$

In an aggregate decomposition,[35] we can consider the term $\sum_{j=1}^{n}\beta_j^{R=1,c}\{E[V_j|R=1,c] - E[V_j|R=0,c]\}$ to comprise the 'explained portion' because it captures the portion of the disparity that is statistically attributable to the fact that the means of explanatory variables differs across racial groups within levels of C. We can consider the terms $(\beta_0^{R=1,c} - \beta_0^{R=0,c})$ and $\sum_{j=1}^{n}(\beta_j^{R=1,c} - \beta_j^{R=0,c})E[V_j|R=0,c]$ and $\sum(\beta_{c'}^{R=1,c} - \beta_{c'}^{R=0,c})c'$ to comprise the 'unexplained portion' because it captures the portion of the disparity that is statistically explained by the fact that associations between the explanatory variables and mean log-wages, and also the associations between the conditioning variables C and mean log-wages, differ by race. It follows then, that if there is no

statistical association between race and the covariates C, the conditional decomposition has the same form as the marginal decomposition except that its components are specific to the levels of the conditioning variables C=c. We could also interpret the components for each explanatory variable $V_j$ in a detailed decomposition[35] as we did so in the marginal decomposition but again, these interpretations would pertain to a specific level of the conditioning variables C=c.

To the best of our knowledge, we have not seen such conditional forms of the Oaxaca-Blinder decomposition considered in the economics or epidemiology literature. While this extension is relatively minor, it has important implications when it comes to the causal interpretation of the decompositions. As described below if an explanatory variable $V_j$ is exchangeable given the conditioning variables C and other explanatory variables that temporally precede $V_j$, then this permits causal inference where the 'explained' portion represents the disparity reduction under an intervention to equalize the explanatory variables $V_j$, and the 'unexplained' portion represents the corresponding residual disparity. We outline this for Propositions 1-4 below and provide all supporting proofs in the eAppendix. The causal interpretations given here are thus with respect to the explanatory variables V, rather than to hypothetical interventions on race itself as per other literature.[35]

Propositions 1-4 expressed as causal implementations of the Oaxaca-Blinder decomposition

Suppose now that we fit three sets of regressions:

Set 1:
$E[Y|R=1,x,c] = \omega_0 + \omega_1 x + \omega_3' c$
$E[Y|R=0,x,c] = \pi_0 + \pi_1 x + \pi_3' c$

Set 2:
$E[Y|R=1,m,x,c] = \alpha_0 + \alpha_1 x + \alpha_2 m + \alpha_3' c$
$E[Y|R=0,m,x,c] = \beta_0 + \beta_1 x + \beta_2 m + \beta_3' c$

Set 3:
$E[Y|r,x,m,c] = \theta_0 + \theta_1 r + \theta_2 x + \theta_3 m + \theta_4 rx + \theta_5 rm + \theta_6' c$
$E[Y|r,x,c] = \gamma_0 + \gamma_1 r + \gamma_2 x + \gamma_4 rx + \gamma_6' c$
$E[Y|r,c] = \phi_0 + \phi_1 r + \phi_6' c$

Suppose further that, for simplicity but not out of necessity, we assume no statistical interactions between race R and covariates C for the mean outcome log-wages i.e. $\omega_3 = \pi_3$ and $\alpha_3 = \beta_3$, such that the models of set 1 are equivalent to the second model in set 3, and the models in set 2 are equivalent to the first model in set 3. The models are equivalent in the sense that they allow for heterogeneous effects of childhood SES X and test scores M across race R. Note that all of our arguments assume no interactions between race and conditioning covariates C, but this was only done to simplify the proofs in the eAppendix.

*Goal of Proposition 1: equalize childhood SES across race given covariates i.e. standardization*

We can carry out an aggregate Oaxaca-Blinder decomposition to understand the extent to which differences in childhood SES X statistically explain the racial disparity within levels of gender and age C. With the models in set 1, the unexplained portion equals $(\omega_0 - \pi_0) + (\omega_1 - \pi_1)E[X|R=0,c]$, and the explained portion equals $\omega_1\{E[X|R=1,c] - E[X|R=0,c]\}$. Now, consider the linear models of set 3 and assume that the effect of childhood SES X on log-wages is unconfounded given covariates

gender and age C=c holds (assumption A1). Under Proposition 1, an intervention to set the distribution of childhood SES X among blacks according to its distribution among whites with covariates C=c, we have that the residual disparity equals: $\gamma_1 + \gamma_4 E[X|R=0,c]$, and the disparity reduction equals: $(\gamma_2 + \gamma_4) \{E[X|R=1,c] - E[X|R=0,c]\}$. We show in the eAppendix that the unexplained portion and the residual disparity are equal, and likewise the explained portion and the disparity reduced are equal.

*Goal of Proposition 2: equalize test scores across race given childhood SES and covariates i.e. mediation-analysis*

We can carry out an aggregate Oaxaca-Blinder decomposition to understand the extent to which differences in test scores M statistically explain the racial disparity within levels of childhood SES X, gender and age C. With the models in set 2, the unexplained portion equals: $(\alpha_0-\beta_0) + (\alpha_1-\beta_1)x + (\alpha_2-\beta_2)E[M|R=0,x,c]$, and the explained portion equals: $\alpha_2\{E[M|R=1,x,c] - E[M|R=0,x,c]\}$. Now, consider the linear models of set 3 and assume that the effect of test scores M on log-wages is unconfounded given childhood SES X and covariates gender and age C=c holds (i.e. assumption A2). Under Proposition 2, an intervention to set the distribution of test scores M among blacks according to its distribution among whites with childhood SES X=x and covariates gender and age C=c, we have that the residual disparity is equal to $\theta_1 + \theta_4 x + \theta_5 E[M|R=0,x,c]$, and the disparity reduction is equal to $(\theta_3 + \theta_5)\{E[M|R=1,x,c] - E[M|R=0,x,c]\}$. We show in the eAppendix that the unexplained portion and the residual disparity are equal, and likewise the explained portion and the disparity reduced are equal.

*Goal of Proposition 3: equalize childhood SES and test scores across race given covariates*

We can carry out an aggregate Oaxaca-Blinder decomposition to understand the extent to which differences in childhood SES X and test scores M statistically explain the racial disparity within levels of covariates gender and age C. With the models in set 2, the unexplained portion equals: $(\alpha_0-\beta_0) + (\alpha_1-\beta_1)E[X|R=0,c] + (\alpha_2-\beta_2)E[M|R=0,c]$, and the explained portion equals: $\alpha_1\{E[X|R=1,c] - E[X|R=0,c]\} + \alpha_2\{E[M|R=1,c] - E[M|R=0,c]\}$. Now, consider the linear models of set 3 and assumptions A1 and A2. Under Proposition 3, an intervention to set the distribution of childhood SES X and test scores M among blacks according to their distribution among whites with covariates C=c, we have that the residual disparity is equal to $\theta_1 + \theta_4 E[X|R=0,c] + \theta_5 E[M|R=0,c]$, and the disparity reduction is equal to $(\theta_2 + \theta_4) \{E[X|R=1,c] - E[X|R=0,c]\} + (\theta_3 + \theta_5) \{E[M|R=1,c] - E[M|R=0,c]\}$. We show in the eAppendix that the unexplained portion and the residual disparity are equal, and likewise the explained portion and the disparity reduced are equal.

*Goal of Proposition 4: equalize test scores across race given covariates*

We can carry out a detailed Oaxaca-Blinder decomposition to understand the extent to which differences in childhood SES X, and also differences in test scores M, each statistically explain, independent of each other, the racial disparity within levels of gender and age C. With the models in set 2, the part of the unexplained portion due to racial differences in the association between childhood SES X and log-wages equals $(\alpha_1-\beta_1)E[X|R=0,c]$; the part of the unexplained portion due to racial differences in the association between test scores M and log-wages equals $(\alpha_2-\beta_2)E[M|R=0,c]$; the part of the explained portion due to racial differences in the distribution of childhood SES X (independent of racial differences in test scores M) equals $\alpha_1\{E[X|R=1,c] - E[X|R=0,c]\}$; the part of the explained portion due to racial differences in the distribution of test scores M (independent of racial differences in childhood SES X) equals $\alpha_2\{E[M|R=1,c] - E[M|R=0,c]\}$. Now, consider the linear models of set 3 and assumption A2. Under Proposition 4, an intervention to set the distribution of test scores M among blacks according to their distribution among whites with covariates gender and age C=c, we have that the residual disparity is equal to $\theta_1 + \theta_2\{E[X|R=1,c]-E[X|R=0,c]\} + \theta_4 E[X|R=1,c] + \theta_5 E[M|R=0,c]$, and the disparity reduction is equal to $(\theta_3 + \theta_5)\{E[M|R=1,c] - E[M|R=0,c]\}$. We show in the eAppendix that the portion explained independently by test scores M and the disparity reduction are equal, and that the sum of the entire unexplained portion and the portion independently explain by childhood SES X equals the disparity reduced.

*A further note about causal interpretation under the detailed decomposition*

Note that the detailed decomposition interprets $\alpha_1\{E[X|R=1,c] - E[X|R=0,c]\}$ as the portion of the disparity in log-wages statistically explained by racial differences in the mean of childhood SES X given covariates C gender and age (independent of racial differences in test scores M). However, this does not in general equal the disparity reduction under Proposition 1 i.e. what would occur under equalizing the distribution of childhood SES X across race given covariates C. Only when the effect of childhood SES X on log-wages is not mediated by test scores M, such that $\omega_1=\alpha_1$, would this interpretation apply. Otherwise it is not clear what the causal interpretation is for a detailed decomposition regarding childhood SES X in the models from set 2.

| | Successive linear models for Y $E[Y\|r,x,m,c]$ $= \theta_0 + \theta_1 r + \theta_2 x + \theta_3 m + \theta_4'c$ $E[Y\|r,x,c]$ $= \gamma_0 + \gamma_1 r + \gamma_2 x + \gamma_4'c$ $E[Y\|r,c]$ $= \phi_0 + \phi_1 r + \phi_4'c$ | Linear models for Y, M, X $E[Y\|r,x,m,c]$ $= \theta_0 + \theta_1 r + \theta_2 x + \theta_3 m + \theta_4'c$ $E[M\|r,x,c]$ $= \beta_0 + \beta_1 r + \beta_2 x + \beta_3'c$ $E[X\|r,c]$ $= \alpha_0 + \alpha_1 r + \alpha_2'c$ |
|---|---|---|
| **Table 4. Results under parametric regression models for a continuous outcome Y** | | |
| Proposition 1 | | |
|   Residual disparity[a] | $\gamma_1$ | $\theta_1 + \beta_1\theta_3$ |
|   Disparity reduction[b] | $\phi_1 - \gamma_1$ | $\alpha_1\theta_2 + \alpha_1\beta_2\theta_3$ |
| Proposition 2 | | |
|   Residual disparity[a] | $\theta_1$ | $\theta_1$ |
|   Disparity reduction[b] | $\gamma_1 - \theta_1$ | $\beta_1\theta_3$ |
| Proposition 3 | | |
|   Residual disparity[a] | $\theta_1$ | $\theta_1$ |
|   Disparity reduction[b] | $\phi_1 - \theta_1$ | $\alpha_1\theta_2 + \beta_1\theta_3 + \alpha_1\beta_2\theta_3$ |
| Proposition 4 | | |
|   Residual disparity[a] | $\theta_1 + (\theta_2/\gamma_2)(\phi_1 - \gamma_1)$ | $\theta_1 + \alpha_1\theta_2$ |
|   Disparity reduction[b] | $(\gamma_1 - \theta_1) + (1-\theta_2/\gamma_2)(\phi_1 - \gamma_1)$ | $\beta_1\theta_3 + \alpha_1\beta_2\theta_3$ |
| [a] $\mu - E[Y\|R=0,c]$ | | |
| [b] $E[Y\|R=1,c] - \mu$ | | |
| where $\mu$ equals the mean counterfactual outcome for group R=1 under the proposed intervention | | |

| | Successive logistic models for Y<br>Logit P[Y\|r,x,m,c]<br>$= \theta_0 + \theta_1 r + \theta_2 x + \theta_3 m + \theta_4' c$<br>Logit P[Y\|r,x,c]<br>$= \gamma_0 + \gamma_1 r + \gamma_2 x + \gamma_4' c$<br>Logit P[Y\|r,c]<br>$= \phi_0 + \phi_1 r + \phi_4' c$ | Models for Y, M, X<br>Logit P[Y\|r,x,m,c]<br>$= \theta_0 + \theta_1 r + \theta_2 x + \theta_3 m + \theta_4' c$<br>E[M\|r,x,c]<br>$= \beta_0 + \beta_1 r + \beta_2 x + \beta_3' c$<br>E[X\|r,c]<br>$= \alpha_0 + \alpha_1 r + \alpha_2' c$ |
|---|---|---|
| **Proposition 1** | | |
| Residual disparity[a] | $\exp\{\gamma_1\}$ | $\exp\{\theta_1 + \beta_1 \theta_3\}$ |
| Disparity reduction[b] | $\exp\{\phi_1 - \gamma_1\}$ | $\exp\{\alpha_1 \theta_2 + \alpha_1 \beta_2 \theta_3\}$ |
| **Proposition 2** | | |
| Residual disparity[a] | $\exp\{\theta_1\}$ | $\exp\{\theta_1\}$ |
| Disparity reduction[b] | $\exp\{\gamma_1 - \theta_1\}$ | $\exp\{\beta_1 \theta_3\}$ |
| **Proposition 3** | | |
| Residual disparity[a] | $\exp\{\theta_1\}$ | $\exp\{\theta_1\}$ |
| Disparity reduction[b] | $\exp\{\phi_1 - \theta_1\}$ | $\exp\{\alpha_1 \theta_2 + \beta_1 \theta_3 + \alpha_1 \beta_2 \theta_3\}$ |
| **Proposition 4** | | |
| Residual disparity[a] | $\exp\{\theta_1 + (\theta_2/\gamma_2)(\phi_1 - \gamma_1)\}$ | $\exp\{\theta_1 + \alpha_1 \theta_2\}$ |
| Disparity reduction[b] | $\exp\{(\gamma_1 - \theta_1) + (1-\theta_2/\gamma_2)(\phi_1 - \gamma_1)\}$ | $\exp\{\beta_1 \theta_3 + \alpha_1 \beta_2 \theta_3\}$ |

Table 5. Results under parametric regression models for a rare binary outcome Y

[a] $\mu$ -E[Y\|R=0,c]
[b] E[Y\|R=1,c]- $\mu$
where $\mu$ equals the mean counterfactual outcome for group R=1 under the proposed intervention

| Appendix Table 1. Characteristics of males in the 1997 National Survey of American Youth Analytic Cohort, mean (standard error) | | |
|---|---|---|
| | White (n=2,413) | Black (n=1,169) |
| Age | 24.6 (1.5) | 24.5 (1.5) |
| | | |
| Adult outcomes | | |
| Wage (dollars / hr) | 21.3 (13.6) | 17.3 (10.1) |
| Unemployed (%) | 6.9 (2.5) | 17.5 (3.8) |
| Incarceration, ever (%) | 8.2 (2.7) | 16.4 (3.7) |
| | | |
| Educational Attainment | | |
| Armed Forces Qualifying Test (AFQT) | 0.35 (0.98) | -0.66 (0.78) |
| Total years education | 12.7 (1.8) | 11.9 (1.4) |
| | | |
| Measures of childhood SES | | |
| Mother's highest grade level | 13.5 (2.5) | 12.5 (2.1) |
| Parental net worth in childhood | $137,933 ($160,945) | $35,994 ($67,260) |
| Household Income in childhood | $59,506 ($46,673) | $30,262 ($29,051) |
| | | |
| Proportion missing (%) | | |
| Missing AFQT | 17.7 (38.2) | 24.8 (43.2) |
| Missing total years of education | 18.1 (38.5) | 16.9 (37.5) |
| Missing mother's highest grade level | 8.1 (27.3) | 14.6 (35.4) |
| Missing parental net worth in childhood | 25.4 (43.6) | 28.0 (44.9) |
| Missing household income in childhood | 22.7 (41.9) | 31.1 (46.3) |

| | Proposition 1 | Proposition 2 | Proposition 3 | Proposition 4 | Re-analysis of Fryer |
|---|---|---|---|---|---|
| | <u>Intervene to equalize the distribution of childhood SES measures across race but not AFQT scores</u> | <u>Intervene to equalize the distribution of AFQT scores across race within levels of childhood SES</u> | <u>Intervene to equalize the distribution of AFQT scores and childhood SES measures across race</u> | <u>Intervene to equalize the distribution of AFQT scores across race but not childhood SES measures</u> | <u>Statistically equalize the distribution of AFQT scores across race without control for childhood SES</u> |
| Log wages | | | | | |
|   Initial disparity | -0.19 (0.02) | -0.14 (0.02) | -0.19 (0.02) | -0.19 (0.02) | -0.19 (0.02) |
|   Residual disparity | -0.14 (0.02) | -0.10 (0.03) | -0.10 (0.03) | -0.13 (0.03) | -0.12 (0.03) |
|   % reduction | 25 | 34 | 51 | 32 | 38 |
| | | | | | |
| Incarceration | | | | | |
|   Initial disparity | 2.12 (1.12) | 1.65 (1.13) | 2.12 (1.12) | 2.12 (1.12) | 2.12 (1.12) |
|   Residual disparity | 1.65 (1.13) | 1.22 (1.13) | 1.22 (1.13) | 1.43 (1.13) | 1.39 (1.13) |
|   % reduction | 54 | 34 | 18 | 36 | 32 |
| | | | | | |
| Unemployment | | | | | |
|   Initial disparity | 2.86 (1.15) | 2.39 (1.16) | 2.86 (1.15) | 2.86 (1.15) | 2.86 (1.15) |
|   Residual disparity | 2.39 (1.16) | 1.95 (1.17) | 1.95 (1.17) | 2.21 (1.17) | 2.12 (1.16) |
|   % reduction | 26 | 31 | 49 | 35 | 40 |

Appendix Table 2. Estimates of residual disparities and disparity reductions in adult outcomes under hypothetical intervention strategies on childhood SES measures and/or Armed Forces Qualifying Test scores in the 1997 NLSY Cohort[2]

[2]The analytic sample size was 3279 for wages, 3294 for unemployment, and 4599 for incarceration. All models included mutually exclusive dummy variables for Hispanic ethnicity and mixed race.

| | Proposition 1 | Proposition 2 | Proposition 3 | Proposition 4 | Re-analysis of Fryer |
|---|---|---|---|---|---|
| | Intervene to equalize the distribution of childhood SES measures across race but not total years of education | Intervene to equalize the distribution of total years of education across race within levels of childhood SES | Intervene to equalize the distribution of total years of education and childhood SES measures across race | Intervene to equalize the distribution of total years of education across race but not childhood SES measures | Statistically equalize the distribution of total years of education across race without control for childhood SES |
| **Log wages** | | | | | |
| Initial disparity | -0.19 (0.02) | -0.14 (0.02) | -0.19 (0.02) | -0.19 (0.02) | -0.19 (0.02) |
| Residual disparity | -0.14 (0.02) | -0.13 (0.02) | -0.13 (0.02) | -0.16 (0.03) | -0.15 (0.02) |
| % reduction | 25 | 11 | 33 | 19 | 21 |
| **Incarceration** | | | | | |
| Initial disparity | 2.22 (1.12) | 1.66 (1.14) | 2.22 (1.12) | 2.22 (1.12) | 2.22 (1.12) |
| Residual disparity | 1.66 (1.14) | 1.50 (1.14) | 1.50 (1.14) | 1.74 (1.42) | 1.69 (1.13) |
| % reduction | 46 | 24 | 59 | 41 | 43 |
| **Unemployment** | | | | | |
| Initial disparity | 2.86 (1.13) | 2.39 (1.15) | 2.86 (1.13) | 2.86 (1.13) | 2.86 (1.13) |
| Residual disparity | 2.39 (1.15) | 2.32 (1.15) | 2.32 (1.15) | 2.64 (1.58) | 2.53 (1.14) |
| % reduction | 26 | 6 | 30 | 15 | 18 |

Appendix Table 3. Estimates of residual disparities and disparity reductions in adult outcomes under hypothetical intervention strategies on childhood SES measures and/or total years of education in the 1997 NLSY Cohort[2]

[2] The analytic sample size was 3279 for wages, 3294 for unemployment, and 4599 for incarceration. All models included mutually exclusive dummy variables for Hispanic ethnicity and mixed race.

## Non-parametric results in the presence of time-dependent confounding

Consider a comparison of two race/ethnicity groups and let R denote a binary variable indicating race. Let X be a set of characteristics at birth or early childhood that are potentially manipulable (e.g. early SES measures), let M be one or more characteristics later in life or in adulthood that are potentially manipulable (e.g. educational attainment or adult SES), let Y be some outcome of interest and let C be some other set of covariates at birth (e.g. gender, year-of-birth/age). The overall disparity measure within strata of covariates C (gender and age) would then be $E[Y|R=1,c]-E[Y|R=0,c]$. Unless noted otherwise, we will consider X to be a single measure of characteristics at birth. Suppose that there is a variable L, that may be affected by C,R,X and that affects both M and Y so that it is a confounder of the relationship between M and Y.

Let Y(x) be the value of the outcome that would have been observed for an individual had X been set to x. Likewise let Y(m) be the value of the outcome that would have been observed for an individual had M been set to m. Finally let Y(x,m) be the value of the outcome that would have been observed for an individual had X been set to x.

Unless otherwise noted we will assume:
A1: The effect of X on the outcome Y is unconfounded given (R,C)
A2: The effect of M on the outcome Y is unconfounded given (R,C,X)
A3: The effect of M on the outcome Y is unconfounded given (R,C,X,L)

Proposition 5. Under (A3), the disparity that would remain if the distribution of M for black persons (R=1) with X=x and covariates C=c were set equal to its distribution for white persons (R=0) with X=x and C=c would be:
$\mu_{m|x} - E[Y|R=0,x,c]$
and the amount the disparity is reduced would be:
$E[Y|R=1,x,c] - \mu_{m|x}$
where $\mu_{m|x} = \Sigma_{m,l|x} E[Y|R=1,x,m,c,l] P(l|R=1,x,c) P(m|R=0,x,c)$.

Proposition 6. Under (A1) and (A3), the disparity that would remain if the distribution of (X,M) for black persons (R=1) with covariates C=c were set equal to its distribution for white persons (R=0) with C=c would be:
$\mu_{xm} - E[Y|R=0,c]$
and the amount the disparity is reduced would be:
$E[Y|R=1,c] - \mu_{xm}$
where $\mu_{xm} = \Sigma_{x,m,l} E[Y|R=1,x,m,c,l] P(l|R=1,x,c) P(m|R=0,x,c) P(x|R=0,c)$.

Proposition 7. Under (A3), the disparity that would remain if the distribution of M for black persons (R=1) with covariates C=c were set equal to its distribution for white persons (R=0) with C=c would be:
$\mu_m - E[Y|R=0,c]$
and the amount the disparity is reduced would be:
$E[Y|R=1,c] - \mu_m$
where $\mu_m = \Sigma_{x,m,l} E[Y|R=1,x,m,c,l] P(l|R=1,x,c) P(m|R=0,c) P(x|R=1,c)$.

## Results for proportion of the disparity reduced

Let D equal the total disparity measured on the difference scale $E[Y|R=1,c]-E[Y|R=0,c]$
Let D* equal the residual disparity measured on the difference scale $\mu-E[Y|R=0,c]$
Let R equal the total disparity measured on the relative scale $E[Y|R=1,c]/E[Y|R=0,c]$
Let R* equal the residual disparity measured on the relative scale $\mu/E[Y|R=0,c]$

<u>Using additive disparity measures</u>

Proportion of disparity remaining = D*/D
Proportion of disparity reduced = (D-D*)/D

<u>Using relative disparity measures</u>

Proportion of disparity remaining = (R*-1)/ (R-1)
Proportion of disparity reduced = (R-R*) / (R-1)

## Results for successive linear models given measures of childhood characteristics, $X_1$, $X_2$, $X_3$

Consider the following models:
$E[Y|r,x_1,x_2,x_3,m,c] = \theta_0 + \theta_1 r + \theta_2 x_1 + \theta_3 x_2 + \theta_4 x_3 + \theta_5 m + \theta_6' c$
$E[Y|r,x_1,x_2,x_3,c] = \delta_0 + \delta_1 r + \delta_2 x_1 + \delta_3 x_2 + \delta_4 x_3 + \delta_6' c$
$E[Y|r,x_1,x_2,c] = \eta_0 + \eta_1 r + \eta_2 x_1 + \eta_3 x_2 + \eta_6' c$
$E[Y|r, x_1,c] = \gamma_0 + \gamma_1 r + \gamma_2 x_1 + \gamma_6' c$
$E[Y|r,c] = \phi_0 + \phi_1 r + \phi_6' c$

In Proposition 1 we have:
The residual disparity is:      $\mu_{x1,x2,x3} - E[Y|R=0,c] = \delta_1$
The disparity reduction is:      $E[Y|R=1,c] - \mu_{x1,x2,x3} = \phi_1 - \delta_1$

In Proposition 2 we have:
The residual disparity is:      $\mu_{m|x1,x2,x3} - E[Y|R=0,x_1,x_2,x_3,c] = \theta_1$
The disparity reduction is:      $E[Y|R=1,x_1,x_2,x_3,c] - \mu_{m|x1,x2,x3} = \delta_1 - \theta_1$

In Proposition 3 we have:
The residual disparity is:      $\mu_{x1,x2,x3,m} - E[Y|R=0,c] = \theta_1$
The disparity reduction is:      $E[Y|R=1,c] - \mu_{x1,x2,x3,m} = \phi_1 - \theta_1$

In Proposition 4 we have:
The residual disparity is:
$\mu_m - E[Y|R=0,c]$
$= \theta_1$
$+ \theta_4/\delta_4 (\eta_1 - \delta_1)$
$+ \{\theta_3/\eta_3 + \theta_4/\delta_4 (1 - \delta_3/\eta_3)\}(\gamma_1 - \eta_1)$
$+ \{\theta_2/\gamma_2 + \theta_3/\eta_3(1-\eta_2/\gamma_2) + \theta_4/\delta_4\{(\eta_2 - \delta_2)/\gamma_2 + (1 - \delta_3/\eta_3)(1-\eta_2/\delta_2)\}\}(\phi_1 - \gamma_1)$
The disparity reduction is:
$E[Y|R=0,c] - \mu_m$
$= (\delta_1 - \theta_1)$

$+ (1-\theta_4/\delta_4)(\eta_1 - \delta_1)$
$+ \{(\delta_3 - \theta_3)/\eta_3 + (1 - \theta_4/\delta_4)(1 - \delta_3/\eta_3)\}(\gamma_1 - \eta_1)$
$+ \{(\delta_2 - \theta_2)/\gamma_2 + (\delta_3 - \theta_3)/\eta_3(1-\eta_2/\gamma_2) + (1-\theta_4/\delta_4)\{(\eta_2 - \delta_2)/\gamma_2 + (1- \delta_3/\eta_3)(1-\eta_2/\delta_2)\}\}(\phi_1 - \gamma_1)$

## Proofs

Our assumptions are:
A1: The effect of X on the outcome Y is unconfounded given (R,C)
A2: The effect of M on the outcome Y is unconfounded given (R,C,X)

Formally these are:
A1: $E[Y(x)|R=r,c] = E[Y(x)|R=r,x,c]$
A1': $E[Y(x,m)|R=r,c] = E[Y(x,m)|R=r,x,c]$
A2: $E[Y(m)|R=r,x,c] = E[Y(m)|R=r,x,m,c]$
A2': $E[Y(x,m)|R=r,x,c] = E[Y(x,m)|R=r,x,m,c]$

Non-parametric formulae in the absence of time-dependent confounding

Recall Proposition 1 (VanderWeele and Robinson, 2014). The disparity that would remain if the childhood distribution of X for black persons (R=1) with covariates C=c were set equal to its distribution for white persons (R=0) with C=c would be:
$\mu_x - E[Y|R=0,c]$
and the amount the disparity is reduced would be:
$E[Y|R=1,c] - \mu_x$
where $\mu_x = \Sigma_x E[Y|R=1,x,c]P(x|R=0,c)$.

Proof of Proposition 1: Let $G_{x|c}$ denote a random draw of the distribution of X among those with R=0,C=c i.e. from $P(x|R=0,c)$. If the distribution of X for black persons (R=1) with covariates C=c were set equal to its distribution for white persons (R=0) the average outcome would be:
$E[Y(x=G_{x|c})|R=1,c]$
$= \Sigma_x E[Y(x)|R=1,c, G_{x|c}=x]P(G_{x|c}=x | R=1,c)$
$= \Sigma_x E[Y(x)|R=1,c] P(x|R=0,c)$
$= \Sigma_x E[Y(x)|R=1,x,c] P(x|R=0,c)$ by (A1)
$= \Sigma_x E[Y|R=1,x,c] P(x|R=0,c)$.
From this the result follows.

Recall Proposition 2 (VanderWeele and Robinson, 2014). The disparity that would remain if the distribution of M for black persons (R=1) with covariates C=c and X=x were set equal to its distribution for white persons (R=0) with C=c and X=x would be:
$\mu_{m|x} - E[Y|R=0,x,c]$
and the amount the disparity is reduced would be:
$E[Y|R=1,x,c] - \mu_{m|x}$
where $\mu_{m|x} = \Sigma_m E[Y|R=1,x,m,c]P(m|R=0,x,c)$.

Proof of Proposition 2. Let $G_{m|x,c}$ denote a random draw of the distribution of M among those with R=0,C=c,X=x i.e. from $P(m|R=0,x,c)$. If the distribution of M for black persons (R=1) with covariates C=c and X=x were set equal to its distribution for white persons (R=0) with covariates C=c and X=x would be:

$E[Y(m)=G_{m|x,c})|R=1,x,c]$
$= \Sigma_m E[Y(m)|R=1,x,c, G_{m|x,c}=m] P(G_{m|x,c} = m | R=1,x,c)$
$= \Sigma_m E[Y(m)|R=1,x,c] P(m|R=0,x,c)$
$= \Sigma_m E[Y(m)|R=1,x,m,c] P(m|R=0,x,c)$ by (A2)
$= \Sigma_m E[Y|R=1,x,m,c] P(m|R=0,x,c)$ by (A2)
From this the result follows.

Recall Proposition 3. The disparity that would remain if the distribution of (X,M) for black persons (R=1) with covariates C=c were set equal to its distribution for white persons (R=0) with C=c would be:
$\mu_{xm} - E[Y|R=0,c]$
and the amount the disparity is reduced would be:
$E[Y|R=1,c] - \mu_{xm}$
where $\mu_{xm} = \Sigma_{x,m} E[Y|R=1,x,m,c]P(m|R=0,x,c)P(x|R=0,c)$.

Proof of Proposition 3. Let $G_{xm|c}$ denote a random draw of the distribution of (X,M) among those with R=0,C=c i.e. from P(m,x|R=0,c). If the distribution of (X,M) for black persons (R=1) with covariates C=c were set equal to its distribution for white persons (R=0) with covariates C=c would the average outcome would be:
$E[Y((x,m)=G_{xm|c})|R=1,c]$
$= \Sigma_{x,m} E[Y(x,m)|R=1,c, G_{xm|c}=(x,m)] P(G_{xm|c}=(x,m)| R=1,c)$
$= \Sigma_{x,m} E[Y(x,m)|R=1,c] P(m,x|R=0,c)$
$= \Sigma_{x,m} E[Y(x,m)|R=1,x,c] P(m|R=0,x,c)P(x|R=0,c)$ by (A1')
$= \Sigma_{x,m} E[Y(x,m)|R=1,x,m,c] P(m|R=0,x,c)P(x|R=0,c)$ by (A2')
$= \Sigma_{x,m} E[Y|R=1,x,m,c] P(m|R=0,x,c)P(x|R=0,c)$.
From this the result follows.

Recall Proposition 4. The disparity that would remain if the distribution of M for black persons (R=1) with covariates C=c were set equal to its distribution for white persons (R=0) with C=c would be:
$\mu_m - E[Y|R=0,c]$
and the amount the disparity is reduced would be:
$E[Y|R=1,c] - \mu_m$
where $\mu_m = \Sigma_{x,m} E[Y|R=1,x,m,c]P(m|R=0,c)P(x|R=1,c)$.

Proof of Proposition 4. Let $G_{m|c}$ denote a random draw of the distribution of M among those with R=0,C=c i.e. from P(m|R=0,c). If the distribution of M for black persons (R=1) with covariates C=c were set equal to its distribution for white persons (R=0) with covariates C=c the average outcome would be:
$E[Y(m)=G_{m|c})|R=1,c]$
$= \Sigma_m E[Y(m)|R=1,c, G_{m|c}=m] P(G_{m|c} = m | R=1,c)$
$= \Sigma_m E[Y(m)|R=1,c] P(m|R=0,c)$
$= \Sigma_{x,m} E[Y(m)|R=1,x,c] P(x|R=1,c) P(m|R=0,c)$
$= \Sigma_{x,m} E[Y(m)|R=1,x,m,c] P(x|R=1,c) P(m|R=0,c)$ by (A2)
$= \Sigma_{x,m} E[Y|R=1,x,m,c] P(x|R=1,c) P(m|R=0,c)$.
From this the result follows.

Non-parametric formulae in the presence of time-dependent confounding

Suppose now that there is a variable L, that may be affected by C,R,X and that affects both M and Y so that it is a confounder of the relationship between M and Y.

We will assume:
A1: The effect of X on the outcome Y is unconfounded given (R,C)
A3: The effect of M on the outcome Y is unconfounded given (R,C,X,L)

Formally these are:
A1: $E[Y(x)|R=r,c] = E[Y(x)|R=r,x,c]$
A3: $E[Y(m)|R=r,x,c,l] = E[Y(m)|R=r,x,m,c,l]$
A3': $E[Y(x,m)|R=r,x,c,l] = E[Y(x,m)|R=r,x,m,c,l]$

Recall Proposition 5. Under (A3), the disparity that would remain if the distribution of M for black persons (R=1) with X=x and covariates C=c were set equal to its distribution for white persons (R=0) with X=x and C=c would be:
$\mu_{m|x} - E[Y|R=0,x,c]$
and the amount the disparity is reduced would be:
$E[Y|R=1,x,c] - \mu_{m|x}$
where $\mu_{m|x} = \Sigma_{m,l|x} E[Y|R=1,x,m,c,l] P(l|R=1,x,c) P(m|R=0,x,c)$.

Proof of Proposition 5. Let $G_{xm|x,c}$ denote a random draw of the distribution of M among those with R=0,X=x,C=c i.e. from P(m|R=0,x,c). If the distribution of M for black persons (R=1) with childhood SES X=x and covariates C=c were set equal to its distribution for white persons (R=0) with childhood SES X=x and covariates C=c would the average outcome would be:
$E[Y((m)=G_{m|x,c})|R=1,x,c]$
$= \Sigma_m E[Y(m)|R=1,x,c,G_{m|x,c}=(m)] P(G_{m|x,c}=(m)|R=1,x,c)$
$= \Sigma_m E[Y(m)|R=1,x,c] P(m|R=0,x,c)$
$= \Sigma_{m,l} E[Y(m)|R=1,x,c,l] P(l|R=1,x,c) P(m|R=0,x,c)$
$= \Sigma_{m,l} E[Y(m)|R=1,x,m,c,l] P(l|R=1,x,c) P(m|R=0,x,c)$ by (A3)
$= \Sigma_{m,l} E[Y|R=1,x,m,c,l] P(l|R=1,x,c) P(m|R=0,x,c)$.
From this the result follows.

Recall Proposition 6. Under (A1') and (A3'), the disparity that would remain if the distribution of (X,M) for black persons (R=1) with covariates C=c were set equal to its distribution for white persons (R=0) with C=c would be:
$\mu_{xm} - E[Y|R=0,c]$
and the amount the disparity is reduced would be:
$E[Y|R=1,c] - \mu_{xm}$
where $\mu_{xm} = \Sigma_{x,m,l} E[Y|R=1,x,m,c,l] P(l|R=1,x,c) P(m|R=0,x,c) P(x|R=0,c)$.

Proof of Proposition 6. Let $G_{xm|c}$ denote a random draw of the distribution of (X,M) among those with R=0,C=c i.e. from P(m,x|R=0,c). If the distribution of (X,M) for black persons (R=1) with covariates C=c were set equal to its distribution for white persons (R=0) with covariates C=c would the average outcome would be:
$E[Y((x,m)=G_{xm|c})|R=1,c]$
$= \Sigma_{x,m} E[Y(x,m)|R=1,c, G_{xm|c}=(x,m)] P(G_{xm|c}=(x,m)|R=1,c)$

$= \Sigma_{x,m} E[Y(x,m)|R=1,c] P(m,x|R=0,c)$
$= \Sigma_{x,m} E[Y(x,m)|R=1,x,c] P(m|R=0,x,c)P(x|R=0,c)$ by (A1')
$= \Sigma_{x,m,l} E[Y(x,m)|R=1,x,c,l] P(l|R=1,x,c)P(m|R=0,x,c)P(x|R=0,c)$
$= \Sigma_{x,m,l} E[Y(x,m)|R=1,x,m,c,l] P(l|R=1,x,c) P(m|R=0,x,c)P(x|R=0,c)$ by (A3')
$= \Sigma_{x,m,l} E[Y|R=1,x,m,c,l] P(l|R=1,x,c)P(m|R=0,x,c)P(x|R=0,c)$.
From this the result follows.

Recall Proposition 7. Under (A3), the disparity that would remain if the distribution of M for black persons (R=1) with covariates C=c were set equal to its distribution for white persons (R=0) with C=c would be:
$\mu_m - E[Y|R=0,c]$
and the amount the disparity is reduced would be:
$E[Y|R=1,c] - \mu_m$
where $\mu_m = \Sigma_{x,m,l} E[Y|R=1,x,m,c,l]P(l|R=1,x,c)P(m|R=0,c)P(x|R=1,c)$.

Proof of Proposition 7. Let $G_{m|c}$ denote a random draw of the distribution of M among those with R=0,C=c i.e. from $P(m|R=0,c)$. If the distribution of M for black persons (R=1) with covariates C=c were set equal to its distribution for white persons (R=0) with covariates C=c the average outcome would be:
$E[Y(m)=G_{m|c})|R=1,c]$
$= \Sigma_m E[Y(m)|R=1,c, G_{m|c}=m] P(G_{m|c} = m | R=1,c)$
$= \Sigma_m E[Y(m)|R=1,c] P(m|R=0,c)$
$= \Sigma_{x,m} E[Y(m)|R=1,x,c] P(x|R=1,c) P(m|R=0,c)$
$= \Sigma_{x,m,l} E[Y(m)|R=1,x,c,l] P(l|R=1,x,c)P(x|R=1,c) P(m|R=0,c)$
$= \Sigma_{x,m,l} E[Y(m)|R=1,x,m,c,l] P(l|R=1,x,c)P(x|R=1,c) P(m|R=0,c)$ by (A3)
$= \Sigma_{x,m,l} E[Y|R=1,x,m,c,l] P(l|R=1,x,c) P(x|R=1,c) P(m|R=0,c)$.
From this the result follows.

Successive linear models for Y

(under a single measure of X)

Consider the following models:
$E[Y|r,x,m,c] = \theta_0 + \theta_1 r + \theta_2 x + \theta_3 m + \theta_4'c$
$E[Y|r,x,c] = \gamma_0 + \gamma_1 r + \gamma_2 x + \gamma_4'c$
$E[Y|r,c] = \phi_0 + \phi_1 r + \phi_4'c$

The results under the linear models for Propositions 1 and 2 were shown in VanderWeele and Robinson (2014).

The results under linear models for Proposition 3, to set the distribution of childhood SES and test scores (X,M) among black persons to their distribution among white persons, follow since:
$\mu_{xm} = \Sigma_{x,m} E[Y|R=1,x,m,c] P(m|R=0,x,c)P(x|R=0,c)$.
$= \Sigma_{x,m} (\theta_0 + \theta_1 + \theta_2 x + \theta_3 m + \theta_4'c) P(m|R=0,x,c)P(x|R=0,c)$
$= \theta_0 + \theta_1 + \theta_2 E[X|R=0,c] + \theta_3 E[M|R=0,c] + \theta_4'c$
Similarly,
$E[Y|R=0,c] = E[Y|R=0,x,m,c] P(m|R=0,x,c)P(x|R=0,c)$.
$= \theta_0 + \theta_2 E[X|R=0,c] + \theta_3 E[M|R=0,c] + \theta_4'c$
Thus, $\mu_{xm} - E[Y|R=0,c] = \theta_1$

Moreover,
$E[Y|R=1,c] - \mu_{xm} = \{E[Y|R=1,c] - E[Y|R=0,c]\} - \{\mu_{xm} - E[Y|R=0,c]\} = \phi_1 - \theta_1$

The results under linear models for Proposition 4, to set the distribution of test scores M among black persons to its distribution among white persons, follow since:

$\mu_m = \Sigma_{x,m} E[Y|R=1,x,m,c]P(m|R=0,c)P(x|R=1,c)$
$= \Sigma_{x,m} (\theta_0 + \theta_1 + \theta_2 x + \theta_3 m + \theta_4'c)P(m|R=0,c)P(x|R=1,c)$
$= \theta_0 + \theta_1 + \theta_2 E[X|R=1,c] + \theta_3 E[M|R=0,c] + \theta_4'c$
Similarly,
$E[Y|R=0,c] = \Sigma_{x,m} E[Y|R=0,x,m,c] P(m|R=0,x,c)P(x|R=0,c)$
$= \Sigma_{x,m} (\theta_0 + \theta_2 x + \theta_3 m + \theta_4'c)P(m|R=0,x,c)P(x|R=0,c)$
$= \theta_0 + \theta_2 E[X|R=0,c] + \theta_3 E[M|R=0,c] + \theta_4'c$
Thus, $\mu_m - E[Y|R=0,c] = \theta_1 + \theta_2 \{E[X|R=1,c] - E[X|R=0,c]\}$

Note that:
$E[Y|R=1,c] - E[Y|R=0,c] = \phi_1$
Also:
$E[Y|R=1,c] - E[Y|R=0,c]$
$= \Sigma_x E[Y|R=1,x,c]P(x|R=1,c) - \Sigma_x E[Y|R=0,x,c]P(x|R=0,c)$
$= \Sigma_x (\gamma_0 + \gamma_1 + \gamma_2 x + \gamma_4'c)P(x|R=1,c) - \Sigma_x (\gamma_0 + \gamma_2 x + \gamma_4'c)P(x|R=0,c)$
$= \gamma_1 + \gamma_2 \{E[X|R=1,c] - E[X|R=0,c]\}$
Thus: $\phi_1 = \gamma_1 + \gamma_2 \{E[X|R=1,c] - E[X|R=0,c]\}$
And so: $\{E[X|R=1,c] - E[X|R=0,c]\} = (\phi_1 - \gamma_1)/\gamma_2$

Therefore the remaining disparity is:
$\mu_m - E[Y|R=0,c]$
$= \theta_1 + \theta_2 \{E[X|R=1,c] - E[X|R=0,c]\}$
$= \theta_1 + \theta_2 (\phi_1 - \gamma_1)/\gamma_2$
And the disparity reduction is:
$E[X|R=1,c] - \mu_m = \{E[X|R=1,c] - E[X|R=0,c]\} - \{\mu_m - E[X|R=0,c\}$
$= \gamma_1 + \gamma_2 \{E[X|R=1,c] - E[X|R=0,c]\} - \theta_1 - \theta_2 \{E[X|R=1,c] - E[X|R=0,c]\}$
$= (\gamma_1 - \theta_1) + (\theta_2/\gamma_2)(\phi_1 - \gamma_1)$

Successive linear models for Y

(under multiple measures X i.e. $X_1, X_2, X_3$)

Suppose there were three potentially manipulable measures of early life characteristics $X_1$, $X_2$, $X_3$ as used in the motivating example. It can be shown that the proofs and non-parametric results above regarding propositions 1-4 apply replacing X with $X_1, X_2, X_3$ and x with $x_1, x_2, x_3$. Below we provide results under successive linear models for outcome Y, however it can be shown that the results also apply on the logit scale under successive logistic models for a rare binary outcome Y.

Consider the following linear models:
$E[Y|r, x_1, x_2, x_3, m, c] = \theta_0 + \theta_1 r + \theta_2 x_1 + \theta_3 x_2 + \theta_4 x_3 + \theta_5 m + \theta_6'c$
$E[Y|r, x_1, x_2, x_3, c] = \delta_0 + \delta_1 r + \delta_2 x_1 + \delta_3 x_2 + \delta_4 x_3 + \delta_6'c$
$E[Y|r, x_1, x_2, c] = \eta_0 + \eta_1 r + \eta_2 x_1 + \eta_3 x_2 + \eta_6'c$
$E[Y|r, x_1, c] = \gamma_0 + \gamma_1 r + \gamma_2 x_1 + \gamma_6'c$

$E[Y|r,c] = \phi_0 + \phi_1 r + \phi_6'c$

The results under linear models for proposition 1, to set the distribution of childhood SES X among black persons to its distribution among white persons, follow since:

$\mu_{x_1,x_2,x_3} = \Sigma_{x_1,x_2,x_3} E[Y|R=1, x_1, x_2, x_3, c] P(x_1, x_2, x_3 | R=0, c)$
$= \Sigma_{x_1,x_2,x_3} E[Y|R=1, x_1, x_2, x_3, c] P(x_3|R=0, x_1, x_2, c) P(x_2|R=0, x_1, c) P(x_1|R=0, c)$
$= \Sigma_{x_1,x_2,x_3} (\delta_0 + \delta_1 + \delta_2 x_1 + \delta_3 x_2 + \delta_4 x_3 + \delta_6'c) P(x_3|R=0, x_1, x_2, c) P(x_2|R=0, x_1, c) P(x_1|R=0, c)$
$= \delta_0 + \delta_1 + \delta_2 E[X_1|R=0, c] + \delta_3 E[X_2|R=0, c] + \delta_4 E[X_3|R=0, c] + \delta_6'c$
Similarly,
$E[Y|R=0, c] = \Sigma_{x_1,x_2,x_3} E[Y|R=0, x_1, x_2, x_3, c] P(x_1, x_2, x_3 | R=0, c)$
$= \Sigma_{x_1,x_2,x_3} E[Y|R=0, x_1, x_2, x_3, c] P(x_3|R=0, x_1, x_2, c) P(x_2|R=0, x_1, c) P(x_1|R=0, c)$
$= \Sigma_{x_1,x_2,x_3} (\delta_0 + \delta_2 x_1 + \delta_3 x_2 + \delta_4 x_3 + \delta_6'c) P(x_3|R=0, x_1, x_2, c) P(x_2|R=0, x_1, c) P(x_1|R=0, c)$
$= \delta_0 + \delta_1 + \delta_2 E[X_1|R=0, c] + \delta_3 E[X_2|R=0, c] + \delta_4 E[X_3|R=0, c] + \delta_6'c$
Thus,
$\mu_{x_1,x_2,x_3} - E[Y|R=0, c] = \delta_1$
Moreover,
$E[Y|R=1, c] - \mu_{x_1,x_2,x_3} = \{E[Y|R=1, c] - E[Y|R=0, c]\} - \{\mu_{x_1,x_2,x_3} - E[Y|R=0, c]\} = \phi_1 - \delta_1$

The results under linear models for proposition 2, to set the distribution of test scores M among black persons with childhood SES X=x to its distribution among white persons with childhood SES X=x, follow since:

$\mu_{m|x_1,x_2,x_3} = \Sigma_m E[Y|R=1, m, x_1, x_2, x_3, c] P(m|R=0, x_1, x_2, x_3, c)$
$= \Sigma_m (\theta_0 + \theta_1 + \theta_2 x_1 + \theta_3 x_2 + \theta_4 x_3 + \theta_5 m + \theta_6'c) P(m|R=0, x_1, x_2, x_3, c)$
$= \theta_0 + \theta_1 + \theta_2 x_1 + \theta_3 x_2 + \theta_4 x_3 + \theta_2 E[M|R=0, x_1, x_2, x_3, c] + \theta_6'c$
Similarly,
$E[Y|R=0, x_1, x_2, x_3, c] = \Sigma_m E[Y|R=0, m, x_1, x_2, x_3, c] P(m|R=0, x_1, x_2, x_3, c)$
$= \Sigma_m (\theta_0 + \theta_2 x_1 + \theta_3 x_2 + \theta_4 x_3 + \theta_5 m + \theta_6'c) P(m|R=0, x_1, x_2, x_3, c)$
$= \theta_0 + \theta_2 x_1 + \theta_3 x_2 + \theta_4 x_3 + \theta_2 E[M|R=0, x_1, x_2, x_3, c] + \theta_6'c$
Thus,
$\mu_{m|x_1,x_2,x_3} - E[Y|R=0, x_1, x_2, x_3, c] = \theta_1$
Moreover,
$E[Y|R=1, x_1, x_2, x_3, c] - E[Y|R=0, x_1, x_2, x_3, c] = \delta_1$
And so,
$E[Y|R=1, x_1, x_2, x_3, c] - \mu_{m|x_1,x_2,x_3}$
$= \{E[Y|R=1, x_1, x_2, x_3, c] - E[Y|R=0, x_1, x_2, x_3, c]\} - \{\mu_{x_1,x_2,x_3} - E[Y|R=0, c]\} = \delta_1 - \theta_1$

The results under linear models for proposition 3, to set the distribution of childhood SES and test scores (X,M) among black persons to its distribution among white persons, follow since:

$\mu_{x_1,x_2,x_3,m} = \Sigma_{x_1,x_2,x_3,m} E[Y|R=1, x_1, x_2, x_3, m, c] P(m|R=0, x_1, x_2, x_3, c) P(x_1, x_2, x_3 | R=0, c)$
$= \Sigma_{x_1,x_2,x_3,m} E[Y|R=1, x_1, x_2, x_3, m, c] P(m|R=0, x_1, x_2, x_3, c) P(x_1, x_2, x_3 | R=0, c)$
$= \Sigma_{x_1,x_2,x_3,m} E[Y|R=1, m, x_1, x_2, x_3, c] P(m|R=0, x_1, x_2, x_3, c) P(x_3|R=0, x_1, x_2, c) P(x_2|R=0, x_1, c) P(x_1|R=0, c)$
$= \Sigma_{x_1,x_2,x_3,m} (\theta_0 + \theta_2 x_1 + \theta_3 x_2 + \theta_4 x_3 + \theta_5 m + \theta_6'c) P(m|R=0, x_1, x_2, x_3, c) P(x_3|R=0, x_1, x_2, c)$
$P(x_2|R=0, x_1, c) P(x_1|R=0, c)$
$= \theta_0 + \theta_1 + \theta_2 E[X_1|R=0, c] + \theta_3 E[X_2|R=0, c] + \theta_4 E[X_3|R=0, c] + \theta_5 E[M|R=0, c] + \theta_6'c$
Similarly,
$E[Y|R=0, c] = \Sigma_{x_1,x_2,x_3,m} E[Y|R=0, x_1, x_2, x_3, m, c] P(m|R=0, x_1, x_2, x_3, c) P(x_1, x_2, x_3 | R=0, c)$

$= \theta_0 + \theta_2 E[X_1|R=0,c] + \theta_3 E[X_2|R=0,c] + \theta_4 E[X_3|R=0,c] + \theta_5 E[M|R=0,c] + \theta_6'c$

Thus, $\mu_{x1,x2,x3,m} - E[Y|R=0,c] = \theta_1$

Moreover,

$E[Y|R=1,c] - \mu_{x1,x2,x3,m} = \{E[Y|R=1,c] - E[Y|R=0,c]\} - \{\mu_{x1,x2,x3,m} - E[Y|R=0,c]\} = \phi_1 - \theta_1$

The results follow under linear models for proposition 4, to set the distribution of test scores M among black persons to its distribution among white persons, follow since:

$\mu_m = \Sigma_{x1,x2,x3,m} E[Y|R=1, x_1, x_2, x_3, m, c] P(m|R=0,c) P(x_1, x_2, x_3|R=1,c)$
$= \Sigma_{x1,x2,x3,m} E[Y|R=1, x_1, x_2, x_3, m, c] P(m|R=0,c) P(x_1, x_2, x_3|R=1,c)$
$= \Sigma_{x1,x2,x3,m} E[Y|R=1, m, x_1, x_2, x_3, c] P(m|R=0,c) P(x_3|R=1,x_1,x_2,c) P(x_2|R=1,x_1,c) P(x_1|R=1,c)$
$= \Sigma_{x1,x2,x3,m} (\theta_0 + \theta_2 x_1 + \theta_3 x_2 + \theta_4 x_3 + \theta_5 m + \theta_6'c) P(m|R=0,c) P(x_3|R=1,x_1,x_2,c) P(x_2|R=1,x_1,c) P(x_1|R=0,c)$
$= \theta_0 + \theta_1 + \theta_2 E[X_1|R=1,c] + \theta_3 E[X_2|R=1,c] + \theta_4 E[X_3|R=1,c] + \theta_5 E[M|R=0,c] + \theta_6'c$

Similarly,

$E[Y|R=0,c] = \Sigma_{x1,x2,x3,m} E[Y|R=0, x_1, x_2, x_3, m, c] P(m|R=0, x_1, x_2, x_3, c) P(x_1, x_2, x_3|R=0,c)$
$= \theta_0 + \theta_2 E[X_1|R=0,c] + \theta_3 E[X_2|R=0,c] + \theta_4 E[X_3|R=0,c] + \theta_5 E[M|R=0,c] + \theta_6'c$

And so

$\mu_m - E[Y|R=0,c]$
$= \theta_1 + \theta_2\{E[X_1|R=1,c] - E[X_1|R=0,c]\} + \theta_3\{E[X_2|R=1,c] - E[X_2|R=0,c]\} + \theta_4\{E[X_3|R=1,c] - E[X_3|R=0,c]\}$

Note that:

$E[Y|R=1,c] - E[Y|R=0,c] = \phi_1$

Also:

$E[Y|R=1,c] - E[Y|R=0,c]$
$= \Sigma_{x1,x2,x3} E[Y|R=1, x_1, x_2, x_3, c] P(x_1, x_2, x_3|R=1,c) - \Sigma_{x1,x2,x3} E[Y|R=0, x_1, x_2, x_3, c] P(x_1, x_2, x_3|R=0,c)$
$= \Sigma_{x1,x2,x3} (\delta_0 + \delta_1 + \delta_2 x_1 + \delta_3 x_2 + \delta_4 x_3 + \delta_6'c) P(x_3|R=1,x_1,x_2,c) P(x_2|R=1,x_1,c) P(x_1|R=1,c)$
$- \Sigma_{x1,x2,x3} (\delta_0 + \delta_2 x_1 + \delta_3 x_2 + \delta_4 x_3 + \delta_6'c) P(x_3|R=0,x_1,x_2,c) P(x_2|R=0,x_1,c) P(x_1|R=0,c)$
$= \delta_1 + \delta_2\{E[X_1|R=1,c] - E[X_1|R=0,c]\} + \delta_3\{E[X_2|R=1,c] - E[X_2|R=0,c]\} + \delta_4\{E[X_3|R=1,c] - E[X_3|R=0,c]\}$

Also:

$E[Y|R=1,c] - E[Y|R=0,c]$
$= \Sigma_{x1,x2} E[Y|R=1, x_1, x_2, c] P(x_1, x_2|R=1,c) - \Sigma_{x1,x2} E[Y|R=0, x_1, x_2, c] P(x_1, x_2|R=0,c)$
$= \Sigma_{x1,x2} (\eta_0 + \eta_1 + \eta_2 x_1 + \eta_3 x_2 + \eta_6'c) P(x_2|R=1,x_1,c) P(x_1|R=1,c)$
$- \Sigma_{x1,x2} (\eta_0 + \eta_2 x_1 + \eta_3 x_2 + \eta_6'c) P(x_2|R=0,x_1,c) P(x_1|R=0,c)$
$= \eta_1 + \eta_2\{E[X_1|R=1,c] - E[X_1|R=0,c]\} + \eta_3\{E[X_2|R=1,c] - E[X_2|R=0,c]\}$

Also:

$E[Y|R=1,c] - E[Y|R=0,c]$
$= \Sigma_{x1} E[Y|R=1, x_1, c] P(x_1|R=1,c) - \Sigma_{x1} E[Y|R=0, x_1, c] P(x_1|R=0,c)$
$= \Sigma_{x1} (\gamma_0 + \gamma_1 + \gamma_2 x_1 + \gamma_6'c) P(x_1|R=1,c) - \Sigma_{x1} (\eta_0 + \eta_2 x_1 + \eta_6'c) P(x_1|R=0,c)$
$= \gamma_1 + \gamma_2\{E[X_1|R=1,c] - E[X_1|R=0,c]\}$

Thus:

$E[X_1|R=1,c] - E[X_1|R=0,c] = (\phi_1 - \gamma_1)/\gamma_2$
$E[X_2|R=1,c] - E[X_2|R=0,c] = \{(\gamma_1 - \eta_1) + (1-\eta_2/\gamma_2)(\phi_1 - \gamma_1)\}/\eta_3$
$E[X_3|R=1,c] - E[X_3|R=0,c] = \{(\eta_1 - \delta_1) + (\eta_2 - \delta_2)/\gamma_2 (\phi_1 - \gamma_1) + (1 - \delta_3/\eta_3)\{(\gamma_1 - \eta_1) + (1-\eta_2/\delta_2)(\phi_1 - \gamma_1)\}\}/\delta_4$

Thus, the residual disparity
$\mu_m - E[Y|R=0,c]$

$= \theta_1 + \theta_2\{E[X_1|R=1,c]-E[X_1|R=0,c]\} + \theta_3\{E[X_2|R=1,c]-E[X_2|R=0,c]\} + \theta_4\{E[X_3|R=1,c]-E[X_3|R=0,c]\}$
$= \theta_1$
$+ \theta_2/\gamma_2 (\phi_1 - \gamma_1)$
$+ \theta_3/\eta_3\{(\gamma_1 - \eta_1) + (1-\eta_2/\gamma_2)(\phi_1 - \gamma_1)\}$
$+ \theta_4/\delta_4\{(\eta_1 - \delta_1)+(\eta_2 - \delta_2)/\gamma_2 (\phi_1 - \gamma_1)+(1- \delta_3/\eta_3)\{(\gamma_1 - \eta_1)+(1-\eta_2/\delta_2)(\phi_1 - \gamma_1)\}\}$
$= \theta_1$
$+ \theta_4/\delta_4(\eta_1 - \delta_1)$
$+ \{\theta_3/\eta_3 + \theta_4/\delta_4 (1- \delta_3/\eta_3)\}(\gamma_1 - \eta_1)$
$+ \{\theta_2/\gamma_2 + \theta_3/\eta_3(1-\eta_2/\gamma_2) + \theta_4/\delta_4\{(\eta_2 - \delta_2)/\gamma_2 + (1- \delta_3/\eta_3)(1-\eta_2/\delta_2)\}\}(\phi_1 - \gamma_1)$

And the disparity reduced
$E[Y|R=0,c] - \mu_m$
$= \{E[Y|R=1,c]-E[Y|R=0,c]\}-\{\mu_m - E[Y|R=0,c]\}$
$= \delta_1 + \delta_2\{E[X_1|R=1,c]-E[X_1|R=0,c]\} + \delta_3\{E[X_2|R=1,c]-E[X_2|R=0,c]\} + \delta_4\{E[X_3|R=1,c]-E[X_3|R=0,c]\}$
$-\theta_1 - \theta_2\{E[X_1|R=1,c]-E[X_1|R=0,c]\} - \theta_3\{E[X_2|R=1,c]-E[X_2|R=0,c]\} - \theta_4\{E[X_3|R=1,c]-E[X_3|R=0,c]\}$
$= (\delta_1 - \theta_1) + (\delta_2-\theta_2)\{E[X_1|R=1,c]-E[X_1|R=0,c]\} + (\delta_3-\theta_3)\{E[X_2|R=1,c]-E[X_2|R=0,c]\}$
$+ (\delta_4-\theta_4)\{E[X_3|R=1,c]-E[X_3|R=0,c]\}$
$= (\delta_1 - \theta_1)$
$+ (\delta_2 - \theta_2)/\gamma_2 (\phi_1 - \gamma_1)$
$+ (\delta_3 - \theta_3)/\eta_3\{(\gamma_1 - \eta_1) + (1-\eta_2/\gamma_2)(\phi_1 - \gamma_1)\}$
$+ (1-\theta_4/\delta_4)\{(\eta_1 - \delta_1)+(\eta_2 - \delta_2)/\gamma_2 (\phi_1 - \gamma_1)+(1- \delta_3/\eta_3)\{(\gamma_1 - \eta_1)+(1-\eta_2/\delta_2)(\phi_1 - \gamma_1)\}\}$
$= (\delta_1 - \theta_1)$
$+ (1-\theta_4/\delta_4)(\eta_1 - \delta_1)$
$+ \{(\delta_3 - \theta_3)/\eta_3 + (1 - \theta_4/\delta_4)(1- \delta_3/\eta_3)\}(\gamma_1 - \eta_1)$
$+ \{(\delta_2 - \theta_2)/\gamma_2 + (\delta_3 - \theta_3)/\eta_3(1-\eta_2/\gamma_2) + (1-\theta_4/\delta_4)\{(\eta_2 - \delta_2)/\gamma_2 + (1- \delta_3/\eta_3)(1-\eta_2/\delta_2)\}\}(\phi_1 - \gamma_1)$

<u>Linear models for Y, M and X</u>

Consider the following models:
$E[Y|r,x,m,c] = \theta_0 + \theta_1 r + \theta_2 x + \theta_3 m + \theta_4' c$
$E[M|r,x,c] = \beta_0 + \beta_1 r + \beta_2 x + \beta_3' c$
$E[X|r,c] = \alpha_0 + \alpha_1 r + \alpha_2' c$

The results follow under these linear models for Proposition 4, to set the distribution of test scores M among black persons to its distribution among white persons, since:

$\mu_m = \Sigma_{x,m} E[Y|R=1,x,m,c]P(m|R=0,c)P(x|R=1,c)$
$= \Sigma_{x,m} (\theta_0 + \theta_1 + \theta_2 x + \theta_3 m + \theta_4' c)P(m|R=0,c)P(x|R=1,c)$
$= \theta_0 + \theta_1 + \theta_2 E[X|R=1,c] + \theta_3 E[M|R=0,c] + \theta_4' c$

We also have that:
$E[Y|R=1,c] = \Sigma_{x,m} E[Y|R=1,x,m,c] P(m|R=1,x,c)P(x|R=1,c).$
$= \Sigma_{x,m} (\theta_0 + \theta_1 + \theta_2 x + \theta_3 m + \theta_4' c)P(m|R=1,x,c)P(x|R=1,c).$
$= \theta_0 + \theta_1 + \theta_2 E[X|R=1,c] + \theta_3 E[M|R=1,c] + \theta_4' c$
Thus, $E[Y|R=1,c] - \mu_m = \theta_3 \{E[M|R=1,c] - E[M|R=0,c]\}$

Also:
$E[Y|R=0,c] = \Sigma_{x,m} E[Y|R=0,x,m,c] P(m|R=0,x,c)P(x|R=0,c).$

$= \Sigma_{x,m} (\theta_0 + \theta_2 x + \theta_3 m + \theta_4'c)P(m|R=0,x,c)P(x|R=0,c)$.
$= \theta_0 + \theta_2 E[X|R=0,c] + \theta_3 E[M|R=0,c] + \theta_4'c$
Thus, $\mu_m - E[Y|R=0,c] = \theta_1 + \theta_2 \{E[X|R=1,c] - E[X|R=0,c]\}$

Note that:
$E[M|R=1,c] - E[M|R=0,c]$
$= \Sigma_x E[M|R=1,x,c]P(x|R=1,c) - \Sigma_x E[M|R=0,x,c]P(x|R=0,c)$
$= \Sigma_x (\beta_0 + \beta_1 + \beta_2 x + \beta_3'c)P(x|R=1,c) - \Sigma_x (\beta_0 + \beta_2 x + \beta_3'c)P(x|R=0,c)$
$= \beta_1 + \beta_2 \{E[X|R=1,c] - E[X|R=0,c]\}$
Also:
$E[X|R=1,c] - E[X|R=0,c] = \alpha_1$

Thus, the remaining disparity is:
$\mu_m - E[Y|R=0,c] = \theta_1 + \theta_2 \alpha_1$

And the disparity reduction is:
$E[Y|R=1,c] - \mu_m = \theta_3 \{\beta_1 + \beta_2 \alpha_1\}$

<u>Successive logistic models for a rare binary outcome Y</u>

Consider the following models:
Logit $P[Y|r,x,m,c] = \theta_0 + \theta_1 r + \theta_2 x + \theta_3 m + \theta_4'c$
Logit $P[Y|r,x,c] = \gamma_0 + \gamma_1 r + \gamma_2 x + \gamma_4'c$
Logit $P[Y|r,c] = \phi_0 + \phi_1 r + \phi_4'c$

The results under logistic models for Proposition 4, to set the distribution of test scores M among black persons to its distribution among white persons, follow since:

Under the assumption Logit $P[Y|\cdot] \approx \log P[Y|\cdot]$,
Logit $\mu_m$
$\approx \log \{\Sigma_{x,m} P[Y|R=1,x,m,c] P(m|R=0,x,c)P(x|R=1,c)\}$
$= \log \{\Sigma_{x,m} \exp(\theta_0 + \theta_1 + \theta_2 x + \theta_3 m + \theta_4'c)P(m|R=0,x,c)P(x|R=1,c)\}$
$= \log \{\exp(\theta_0 + \theta_1 + \theta_4'c) E[\exp(\theta_2 X)|R=1,c] E[\exp(\theta_3 M)|R=0,c]\}$
$= \theta_0 + \theta_1 + \theta_4'c + \log E[\exp(\theta_2 X)|R=1,c] + \log E[\exp(\theta_3 M)|R=0,c]\}$

Similarly Logit $E[Y|R=1,c]$
$\approx \log \{\Sigma_{x,m} P[Y|R=1,x,m,c] P(m|R=1,x,c)P(x|R=1,c)\}$
$= \log \{\Sigma_{x,m} \exp(\theta_0 + \theta_1 + \theta_2 x + \theta_3 m + \theta_4'c)P(m|R=1,x,c)P(x|R=1,c)\}$
$= \log \{\exp(\theta_0 + \theta_1 + \theta_4'c) E[\exp(\theta_2 X)|R=1,c] E[\exp(\theta_3 M)|R=1,c]\}$
$= \theta_0 + \theta_1 + \theta_4'c + \log E[\exp(\theta_2 X)|R=1,c] + \log E[\exp(\theta_3 M)|R=1,c]$

Similarly Logit $E[Y|R=0,c]$
$\approx \log \{\Sigma_{x,m} P[Y|R=0,x,m,c] P(m|R=0,x,c)P(x|R=0,c)\}$
$= \log \{\Sigma_{x,m} \exp(\theta_0 + \theta_2 x + \theta_3 m + \theta_4'c)P(m|R=0,x,c)P(x|R=0,c)\}$
$= \log \{\exp(\theta_0 + \theta_4'c) E[\exp(\theta_2 X)|R=0,c] E[\exp(\theta_3 M)|R=0,c]\}$
$= \theta_0 + \theta_4'c + \log E[\exp(\theta_2 X)|R=0,c] + \log E[\exp(\theta_3 M)|R=0,c]$

Note that:
Logit $P[Y|R=1,c] -$ Logit $P[Y|R=0,c]$

$\approx \text{Log } P[Y|R=1,c] - \text{Log } P[Y|R=0,c]$
$= \phi_1$

Also note that
Logit $P[Y|R=1,c]$ - Logit $P[Y|R=0,c]$
$\approx \text{Log } \{\Sigma_x P[Y|R=1,x,c] P(x|R=1,c)\} - \text{Log } \{\Sigma_x P[Y|R=0,x,c] P(x|R=0,c)\}$
$= \text{Log } \{\Sigma_x \exp(\gamma_0 + \gamma_1 + \gamma_2 x + \gamma_4'c) P(x|R=1,c)\} - \text{Log } \{\Sigma_x \exp(\gamma_0 + \gamma_2 x + \gamma_4'c) P(x|R=0,c)\}$
$= \text{Log } \{\exp(\gamma_0 + \gamma_1 + \gamma_4'c) E[\exp(\gamma_2 X)|R=1,c]\} - \text{Log } \{\exp(\gamma_0 + \gamma_4'c) E[\exp(\gamma_2 X)|R=1,c]\}$
$= \gamma_1 + \text{Log } E[\exp(\gamma_2 X)|R=1,c] - \text{Log } E[\exp(\gamma_2 X)|R=0,c]$
$= \gamma_1 + \gamma_2 E[X|R=1,c] + \frac{1}{2}(\gamma_2)^2 \sigma_X^2 - \gamma_2 E[X|R=0,c] - \frac{1}{2}(\gamma_2)^2 \sigma_X^2$
$= \gamma_1 + \gamma_2 \{E[X|R=1,c] - E[X|R=0,c]\}$
And so
$\phi_1 = \gamma_1 + \gamma_2 \{E[X|R=1,c] - E[X|R=0,c]\}$
$\{E[X|R=1,c] - E[X|R=0,c]\} = (\phi_1 - \gamma_1)/\gamma_2$

Thus, the remaining disparity is equal to
Logit $\mu_m$ - Logit $E[Y|R=0]$
$= \theta_1 + \log E[\exp(\theta_2 X)|R=1,c] - \log E[\exp(\theta_2 X)|R=0,c]$
$= \theta_1 + \log \{\exp(\theta_2 E[X|R=1,c] + \frac{1}{2}(\theta_2)^2 \sigma_X^2)\} - \log \{\exp(\theta_2 E[X|R=0,c] + \frac{1}{2}(\theta_2)^2 \sigma_X^2)\}$
$= \theta_1 + \theta_2 \{E[X|R=1,c] - E[X|R=0,c]\}$
$= \theta_1 + \theta_2 (\phi_1 - \gamma_1)/\gamma_2$

And the disparity reduction is equal to
Logit $E[Y|R=1]$ - Logit $\mu_m$
$= (\text{Logit } E[Y|R=1] - \text{Logit } E[Y|R=0]) - (\text{Logit } E[Y|R=1] - \text{Logit } \mu_{xm})$
$= \gamma_1 + \gamma_2 \{E[X|R=1,c] - E[X|R=0,c]\} - \theta_1 - \theta_2 \{E[X|R=1,c] - E[X|R=0,c]\}$
$= (\gamma_1 - \theta_1) + (1 - \theta_2/\gamma_2)(\phi_1 - \gamma_1)$

<u>Logistic model for a rare binary outcome Y with linear models for M and X</u>

Consider the following models:
Logit $P[Y|r,x,m,c] = \theta_0 + \theta_1 r + \theta_2 x + \theta_3 m + \theta_4'c$
$E[M|r,x,c] = \beta_0 + \beta_1 r + \beta_2 x + \beta_4'c$
$E[X|r,c] = \alpha_0 + \alpha_1 r + \alpha_4'c$

Assume the outcome is rare and the error term in the model for $E[X|r,c]$ is normally distributed and constant variance $\sigma_x$, and the error term in the model for $E[M|r,x,c]$ is normally distributed with constant variance $\sigma_M$

The results under these models for Proposition 4, to set the distribution of test scores M among black persons to its distribution among white persons, follow since:

Under the assumption Logit $P[Y|\cdot] \approx \log P[Y|\cdot]$, we have that Logit $\mu_{xm}$
$\approx \text{Log } \{\Sigma_{x,m} P[Y|R=1,x,m,c] P(m|R=0,x,c) P(x|R=1,c)\}$
$= \text{Log } \{\Sigma_{x,m} \exp(\theta_0 + \theta_1 + \theta_2 x + \theta_3 m + \theta_4'c) P(m|R=0,x,c) P(x|R=1,c)\}$
$= \text{Log } \{\exp(\theta_0 + \theta_1 + \theta_4'c) E[\exp(\theta_2 X)|R=1,c] E[\exp(\theta_3 M)|R=0,c]\}$
$= \text{Log } \{\exp(\theta_0 + \theta_1 + \theta_4'c) \exp((\theta_2)(\alpha_0 + \alpha_1 + \alpha_4'c) + \frac{1}{2}(\theta_2)^2 \sigma_x^2) E[\exp(\theta_3 M)|R=0,c]\}$
$= \text{Log } \{\exp(\theta_0 + \theta_1 + \theta_4'c) \exp((\theta_2)(\alpha_0 + \alpha_1 + \alpha_4'c) + \frac{1}{2}(\theta_2)^2 \sigma_x^2) \exp((\theta_3) E[M|R=0,c] + \frac{1}{2}(\theta_3)^2 \sigma_M^2)\}$

$= \theta_0 + \theta_1 + \theta_4'c + \theta_2(\alpha_0 + \alpha_1 + \alpha_4'c) + \frac{1}{2}(\theta_2)^2 \sigma_x^2 + (\theta_3)E[M|R=0,c] + \frac{1}{2}(\theta_2)^2 \sigma_M^2$

Similarly Logit $E[Y|R=1,c]$
$\approx \text{Log} \{\Sigma_{x,m} P[Y|R=1,x,m,c] P(m|R=1,x,c)P(x|R=1,c)\}$
$= \text{Log} \{\Sigma_{x,m} \exp(\theta_0 + \theta_1 + \theta_2 x + \theta_3 m + \theta_4'c)P(m|R=1,x,c)P(x|R=1,c)\}$
$= \text{Log} \{\exp(\theta_0 + \theta_1 + \theta_4'c) E[\exp(\theta_2 X)|R=1,c] E[\exp(\theta_3 M)|R=1,c]\}$
$= \text{Log} \{\exp(\theta_0 + \theta_1 + \theta_4'c) \exp((\theta_2)(\alpha_0 + \alpha_1 + \alpha_4'c) + \frac{1}{2}(\theta_2)^2 \sigma_x^2) \exp((\theta_3)E[M|R=1,c] + \frac{1}{2}(\theta_3)^2 \sigma_M^2)\}$
$= \theta_0 + \theta_1 + \theta_4'c + \theta_2(\alpha_0 + \alpha_1 + \alpha_4'c) + \frac{1}{2}(\theta_2)^2 \sigma_x^2 + (\theta_3)E[M|R=1,c] + \frac{1}{2}(\theta_2)^2 \sigma_M^2$

Similarly Logit $E[Y|R=0,c]$
$\approx \text{Log} \{\Sigma_{x,m} P[Y|R=0,x,m,c] P(m|R=0,x,c)P(x|R=0,c)\}$
$= \text{Log} \{\Sigma_{x,m} \exp(\theta_0 + \theta_2 x + \theta_3 m + \theta_4'c)P(m|R=0,x,c)P(x|R=0,c)\}$
$= \text{Log} \{\exp(\theta_0 + \theta_4'c) E[\exp(\theta_2 X)|R=0,c] E[\exp(\theta_3 M)|R=0,c]\}$
$= \text{Log} \{\exp(\theta_0 + \theta_4'c) \exp((\theta_2)(\alpha_0 + \alpha_4'c) + \frac{1}{2}(\theta_2)^2 \sigma_x^2) \exp((\theta_3)E[M|R=0,c] + \frac{1}{2}(\theta_3)^2 \sigma_M^2)\}$
$= \theta_0 + \theta_4'c + \theta_2(\alpha_0 + \alpha_4'c) + \frac{1}{2}(\theta_2)^2 \sigma_x^2 + (\theta_3)E[M|R=0,c] + \frac{1}{2}(\theta_2)^2 \sigma_M^2$

Note that logit $E[M|R=1,c]$ – logit $E[M|R=0,c] \approx$
$= \Sigma_x E[M|R=1,x,c]P(x|R=1,c) - \Sigma_x E[M|R=0,x,c]P(x|R=0,c)$
$= \Sigma_x (\beta_0 + \beta_1 + \beta_2 x + \beta_3'c)P(x|R=1,c) - \Sigma_x (\beta_0 + \beta_2 x + \beta_3'c)P(x|R=0,c)$
$= \beta_1 + \beta_2 \{E[X|R=1,c] - E[X|R=0,c]\}$
$= \beta_1 + \beta_2 \alpha_1$

Thus the disparity reduction $E[Y|R=1,c]/\mu_{xm}$ is
$= \exp(\text{Logit } E[Y|R=1,c] - \text{Logit } \mu_{xm})$
$\approx \exp(\text{Log } E[Y|R=1,c] - \text{Log } \mu_{xm})$
$= \exp(\theta_3\{E[M|R=1,c]-E[M|R=0,c]\})$
$= \exp(\theta_3\{\beta_1 + \beta_2\alpha_1\})$

And the remaining disparity $\mu_{xm}/E[Y|R=0,c]$ is
$= \exp(\text{Logit } \mu_{xm} - \text{Logit } E[Y|R=0,c])$
$\approx \exp(\text{Log } \mu_{xm} - \text{Log } E[Y|R=0,c])$
$= \exp(\theta_1 + \theta_2\alpha_1)$

Oaxaca-Blinder decomposition

Consider the two sets of race-stratified linear models that each can be used to carry out different Oaxaca-Blinder decompositions:

Set 1:
$E[Y|R=1,x,c]=\omega_0 + \omega_1 x + \omega_3'c$
$E[Y|R=0,x,c]=\pi_0 + \pi_1 x + \pi_3'c$

To simplify the formulas we derive, we assume that $\omega_3=\pi_3$. We could allow for $\omega_3 \neq \pi_3$ but this is does not materially affect our proof that propositions 1-4 can be expressed as causal implementations of the Oaxaca-Blinder decomposition.

Set 2:
$E[Y|R=1,m,x,c]=\alpha_0 + \alpha_1 x + \alpha_2 m + \alpha_3'c$

$E[Y|R=0,m,x,c] = \beta_0 + \beta_1 x + \beta_2 m + \beta_3'c$

Consider also successive linear models for Y, this time with interaction terms between R and X and also R and M. (These models could allow for interactions between R and C, and while this would slightly change some of the formulas we derive, this additional complexity does not affect the ability to express propositions 1-4 as causal implementations of the Oaxaca-Blinder decomposition).

Set 3:
$E[Y|r,x,m,c] = \theta_0 + \theta_1 r + \theta_2 x + \theta_3 m + \theta_4 rx + \theta_5 rm + \theta_6'c$
$E[Y|r,x,c] = \gamma_0 + \gamma_1 r + \gamma_2 x + \gamma_4 rx + \gamma_6'c$
$E[Y|r,c] = \phi_0 + \phi_1 r + \phi_6'c$

Again, we could incorporate interaction terms between race and the covariates C in these models, but again, this additional complexity would not affect the ability to express propositions 1-4 as causal implementations of the Oaxaca-Blinder deocmponsition.

For Proposition 1 (i.e. equalize the distribution of childhood SES X across race R), the results under an Oaxaca-Blinder decomposition with models from set 1 equate to results using linear models from set 3 since, under assumption A1:

$\mu_x = \Sigma_x E[Y|R=1,x,c]P(x|R=0,c)$
$= \Sigma_x (\gamma_0 + \gamma_1 + \gamma_2 x + \gamma_4 x + \gamma_6'c)P(x|R=0,c)$
$= \gamma_0 + \gamma_1 + (\gamma_2 + \gamma_4) E[X|R=0,c] + \gamma_6'c$
Similarly,
$E[Y|R=0,c] = E[Y|R=0,x,c]P(x|R=0,c)$
$= \Sigma_x (\gamma_0 + \gamma_2 x + \gamma_6'c)P(x|R=0,c)$
$= \gamma_0 + \gamma_2 E[X|R=0,c] + \theta_6'c$
Thus, $\mu_x - E[Y|R=0,c] = \gamma_1 + \gamma_4 E[X|R=0,c]$
Also,
$E[Y|R=1,c] = E[Y|R=1,x,m,c]P(x|R=1,c)$
$= \Sigma_x (\gamma_0 + \gamma_1 r + \gamma_2 x + \gamma_4 x + \gamma_6'c)P(x|R=0,c)$
$= (\gamma_0 + \gamma_1 + (\gamma_2 + \gamma_4) E[X|R=1,c] + \gamma_6'c$
Thus, $E[Y|R=1,c] - \mu_x = (\gamma_2 + \gamma_4) \{E[X|R=1,c] - E[X|R=0,c]\}$

Note that
$E[Y|R=1,c] = \Sigma_x E[Y|R=1,x,c]P(x|R=1,c)$
$= \Sigma_x (\omega_0 + \omega_1 x + \omega_3'c) P(x|R=1,c)$
$= \omega_0 + \omega_1 E[X|R=1,c] + \omega_3'c$
Similarly,
$E[Y|R=0,c] = \Sigma_x E[Y|R=0,x,c]P(x|R=0,c)$
$= \Sigma_x (\pi_0 + \pi_1 x + \pi_3'c)P(x|R=0,c)$
$= \pi_0 + \pi_1 E[X|R=0,c] + \pi_3'c$

Thus,
$E[Y|R=1,c] - E[Y|R=0,c]$
$= (\omega_0 - \pi_0) + \omega_1 E[X|R=1,c] - \pi_1 E[X|R=0,c]$
$= (\omega_0 - \pi_0) + \omega_1 E[X|R=1,c] - \pi_1 E[X|R=0,c] + \omega_1 E[X|R=0,c] - \omega_1 E[X|R=0,c]$
$= (\omega_0 - \pi_0) + (\omega_1 - \pi_1)E[X|R=0,c] + \omega_1 \{E[X|R=1,c] - E[X|R=0,c]\}$

In an Oaxaca-Blinder decomposition, the terms $(\omega_0 - \pi_0)$ and $(\omega_1 - \pi_1)E[X|R=0,c]$ could be referred to as the "unexplained portion" and the term $\omega_1\{E[X|R=1,c]-E[X|R=0,c]\}$ could be referred to as the "explained" portion, whose sum equals the total disparity $E[Y|R=1,c]-E[Y|R=0,c]$.

Note that by definition, $(\omega_0 - \pi_0) = \gamma_1$, $\pi_1 = \gamma_2$, and $(\omega_1 - \pi_1) = \gamma_4$.

Thus,
$\mu_x - E[Y|R=0,c]$
$= \gamma_1 + \gamma_4 E[X|R=0,c]$
$= (\omega_0 - \pi_0) + (\omega_1 - \pi_1)E[X|R=0,c]$
Also,
$E[Y|R=1,c] - \mu_x$
$= (\gamma_2 + \gamma_4)\{E[X|R=1,c] - E[X|R=0,c]\}$
$= \omega_1\{E[X|R=1,c] - E[X|R=0,c]\}$

Thus, under assumption A1, these quantities can be interpreted as the residual disparity and disparity reduction under an intervention to equalize X alone (proposition 1). Note that if $\gamma_4=0$ such that $\omega_1=\pi_1$ we obtain the results under linear models in the main text.

For Proposition 2 (i.e. equalize the distribution of test scores M across race R within levels of childhood SES X), the results under an Oaxaca-Blinder decomposition with models from set 2 equate to results using linear models from set 3 since under assumption A2:

$\mu_{m|x} = \Sigma_m E[Y|R=1,x,m,c]P(m|R=0,x,c)$
$= \Sigma_x (\theta_0 + \theta_1 + \theta_2 x + \theta_3 m + \theta_4 x + \theta_5 m + \theta_6'c)P(m|R=0,x,c)$
$= \theta_0 + \theta_1 + (\theta_2 + \theta_4) x + (\theta_3 + \theta_5) E[M|R=0,x,c] + \theta_6'c$
Similarly,
$E[Y|R=0,x,c] = \Sigma_m E[Y|R=0,x,m,c]P(m|R=0,x,c)$
$= \Sigma_m (\theta_0 + \theta_2 x + \theta_3 m + \theta_6'c)P(m|R=0,x,c)$
$= \theta_0 + \theta_2 x + \theta_5 E[M|R=0,x,c] + \theta_6'c$
Thus, $\mu_{m|x} - E[Y|R=0,c] = \theta_1 + \theta_4 x + \theta_5 E[M|R=0,x,c]$
Also,
$E[Y|R=1,x,c] = E[Y|R=1,x,m,c]P(x|R=1,c)$
$= \Sigma_m (\theta_0 + \theta_1 + \theta_2 x + \theta_3 m + \theta_4 x + \theta_5 m + \theta_6'c)P(m|R=1,x,c)$
$= \theta_0 + \theta_1 + (\theta_2 + \theta_4) x + (\theta_3 + \theta_5) E[M|R=1,x,c] + \theta_6'c$
Thus, $E[Y|R=1,x,c] - \mu_{m|x} = (\theta_3 + \theta_5)\{E[M|R=1,x,c] - E[M|R=0,x,c]\}$

Note that
$E[Y|R=1,x,c] = \Sigma_m E[Y|R=1,m,x,c]P(m|R=1,x,c)$
$= \Sigma_m (\alpha_0 + \alpha_1 x + \alpha_2 m + \alpha_3'c) P(m|R=1,x,c)$
$= \alpha_0 + \alpha_1 x + \alpha_2 E[M|R=1,x,c] + \alpha_3'c$
Similarly,
$E[Y|R=0,x,c] = \Sigma_m E[Y|R=0,m,x,c]P(m|R=0,x,c)$
$= \Sigma_m (\beta_0 + \beta_1 x + \beta_2 m + \beta_3'c)P(m|R=0,x,c)$
$= \beta_0 + \beta_1 x + \beta_2 E[M|R=0,x,c] + \beta_3'c$
Thus,
$E[Y|R=1,x,c] - E[Y|R=0,x,c]$
$= (\alpha_0 - \beta_0) + \alpha_1 x - \beta_1 x + \alpha_2 E[M|R=1,x,c] - \beta_2 E[M|R=0,x,c]$

$= (\alpha_0-\beta_0) + \alpha_1 x - \beta_1 x + \alpha_2 E[M|R=1,x,c] - \beta_2 E[M|R=0,x,c] + \alpha_2 E[M|R=0,x,c] - \alpha_2 E[M|R=0,x,c]$
$= (\alpha_0-\beta_0) + (\alpha_1-\beta_1)x + (\alpha_2-\beta_2)E[M|R=0,x,c] + \alpha_2\{E[M|R=1,x,c]-E[X|R=0,x,c]\}$

In an Oaxaca-Blinder decomposition, the terms $(\alpha_0-\beta_0)$ and $(\alpha_1-\beta_1)x$ and $(\alpha_2-\beta_2)E[M|R=0,x,c]$ could be referred to as the "unexplained portion given X." The term $\alpha_2\{E[M|R=1,x,c]-E[X|R=0,x,c]\}$ could be referred to as the "explained portion given X," whose sum equals the total disparity within levels of X i.e. $E[Y|R=1,x,c]-E[Y|R=0,x,c]$.

Note that by definition $(\alpha_0-\beta_0)=\theta_1$, $(\alpha_1-\beta_1)=\theta_4$, $\beta_2=\theta_3$, and $(\alpha_2-\beta_2) = \theta_5$.

Thus,
$\mu_{m|x} - E[Y|R=0,c]$
$= \theta_1 + \theta_4 x + \theta_5 E[M|R=0,x,c]$
$= (\alpha_0-\beta_0) + (\alpha_1-\beta_1)x + (\alpha_2-\beta_2)E[M|R=0,x,c]$
Also
$E[Y|R=1,x,c] - \mu_{m|x}$
$= (\theta_3 + \theta_5)\{E[M|R=1,x,c] - E[M|R=0,x,c]\}$
$= \alpha_2\{E[M|R=1,x,c] - E[M|R=0,x,c]\}$

Thus, under assumption A2, these quantities can be interpreted as the residual disparity and disparity reduction under an intervention to equalize M within levels of X (proposition 2). Note that if $\theta_4=0$ such that $\alpha_1=\beta_1$ and $\theta_5=0$ such that $\alpha_2=\beta_2$ we obtain the results under linear models in the main text.

For Proposition 3 (i.e. equalize the distribution of childhood SES X and test scores M across race R), the results under an Oaxaca-Blinder decomposition with models from set 2 equate to results using linear models from set 3 since under assumptions A1' and A2':

$\mu_{xm} = \Sigma_{xm} E[Y|R=1,m,x,c] P(m|R=0,x,c)P(x|R=0,c)$
$= \Sigma_{xm} (\theta_0 + \theta_1 + \theta_2 x + \theta_3 m + \theta_4 x + \theta_5 m + \theta_6'c) P(m|R=0,x,c)P(x|R=0,c)$
$= \theta_0 + \theta_1 + (\theta_2 + \theta_4) E[X|R=0,c] + (\theta_3 + \theta_5) E[M|R=0,c] + \theta_6'c$
Similarly,
$E[Y|R=0,c] = \Sigma_{xm} E[Y|R=0,m,x,c] P(m|R=0,x,c)P(x|R=0,c)$
$= \Sigma_{xm} (\theta_0 + \theta_2 x + \theta_3 m + \theta_6'c) P(m|R=0,x,c)P(x|R=0,c)$
$= \theta_0 + \theta_2 E[X|R=0,c] + \theta_3 E[M|R=0,c] + \theta_6'c$
Thus, $\mu_x - E[Y|R=0,c] = \theta_1 + \theta_4 E[X|R=0,c] + \theta_5 E[M|R=0,c]$
Also,
$E[Y|R=1,c] = E[Y|R=1,x,m,c]P(x|R=1,c)$
$= \Sigma_{xm} (\theta_0 + \theta_1 + \theta_2 x + \theta_3 m + \theta_4 x + \theta_5 m + \theta_6'c) P(m|R=1,x,c)P(x|R=1,c)$
$= \theta_0 + \theta_1 + (\theta_2 + \theta_4) E[X|R=1,c] + (\theta_3 + \theta_5) E[M|R=1,c] + \theta_6'c$
Thus, $E[Y|R=1,c] - \mu_{xm} = (\theta_2 + \theta_4) \{E[X|R=1,c] - E[X|R=0,c]\} + (\theta_3 + \theta_5) \{E[M|R=1,c] - E[M|R=0,c]\}$

Note that
$E[Y|R=1,c] = \Sigma_{xm} E[Y|R=1,m,x,c] P(m|R=1,x,c)P(x|R=1,c)$
$= \Sigma_{xm} (\alpha_0 + \alpha_1 x + \alpha_2 m + \alpha_3'c) P(m|R=1,x,c)P(x|R=1,c)$
$= \alpha_0 + \alpha_1 E[X|R=1,c] + \alpha_2 E[M|R=1,c] + \alpha_3'c$
Similarly we have that:
$E[Y|R=0,c] = \Sigma_{xm} E[Y|R=0,x,c] P(m|R=0,x,c)P(x|R=0,c)$
$= \Sigma_{xm} (\beta_0 + \beta_1 x + \beta_2 m + \beta_3'c) P(m|R=0,x,c)P(x|R=0,c)$

$= \beta_0 + \beta_1 E[X|R=0,c] + \beta_2 E[M|R=0,c] + \beta_3'c$

Thus,

$E[Y|R=1,c] - E[Y|R=0,c]$
$= (\alpha_0 - \beta_0) + \alpha_1 E[X|R=1,c] - \beta_1 E[X|R=0,c] + \alpha_2 E[M|R=1,c] - \beta_2 E[M|R=0,c]$
$= (\alpha_0 - \beta_0) + \alpha_1 E[X|R=1,c] - \beta_1 E[X|R=0,c] + \alpha_2 E[M|R=1,c] - \beta_2 E[M|R=0,c]$
$+ \alpha_1 E[X|R=0,c] - \alpha_1 E[X|R=0,c] + \alpha_2 E[M|R=0,c] - \alpha_2 E[M|R=0,c]$
$= (\alpha_0 - \beta_0) + (\alpha_1 - \beta_1) E[X|R=0,c] + (\alpha_2 - \beta_2) E[M|R=0,c] + \alpha_1 \{E[X|R=1,c] - E[X|R=0,c]\} + \alpha_2 \{E[M|R=1,c] - E[M|R=0,c]\}$

In an Oaxaca-Blinder decomposition, the terms $(\alpha_0 - \beta_0)$ and $(\alpha_1 - \beta_1) E[X|R=0,c]$ and $(\alpha_2 - \beta_2) E[M|R=0,c]$ would be referred to as the "unexplained portion" and the third term $\alpha_1 \{E[X|R=1,c] - E[X|R=0,c]\}$ and fourth term $\alpha_2 \{E[M|R=1,c] - E[M|R=0,c]\}$ would be referred to as the "explained" portion.

Note that by definition, $(\alpha_0 - \beta_0) = \theta_1$, $\beta_1 = \theta_2$, $\beta_2 = \theta_3$, $(\alpha_1 - \beta_1) = \theta_4$, and $(\alpha_2 - \beta_2) = \theta_5$.

Thus,
$\mu_x - E[Y|R=0,c]$
$= \theta_1 + \theta_4 E[X|R=0,c] + \theta_5 E[M|R=0,c]$
$= (\alpha_0 - \beta_0) + (\alpha_1 - \beta_1) E[X|R=0,c] + (\alpha_2 - \beta_2) E[M|R=0,c]$
Also,
$E[Y|R=1,c] - \mu_x$
$= (\theta_2 + \theta_4) \{E[X|R=1,c] - E[X|R=0,c]\} + (\theta_3 + \theta_5) \{E[M|R=1,c] - E[M|R=0,c]\}$
$= \alpha_1 \{E[X|R=1,c] - E[X|R=0,c]\} + \alpha_2 \{E[M|R=1,c] - E[M|R=0,c]\}$

Thus, under assumptions A1' and A2', these quantities can be interpreted as the residual disparity and disparity reduction under an intervention to equalize X and M (proposition 3). Note that if $\theta_4 = 0$ such that $\alpha_1 = \beta_1$ and $\theta_5 = 0$ such that $\alpha_2 = \beta_2$ we obtain the results under linear models in the main text.

For Proposition 4 (i.e. equalize the distribution of test scores M across race R), the results under a detailed Oaxaca-Blinder decomposition, with models from set 2, can be used to obtain results using linear models from set 3, since under assumption A2:

$\mu_m = \Sigma_{xm} E[Y|R=1,m,x,c] P(m|R=0,x,c) P(x|R=1,c)$
$= \Sigma_{xm} (\theta_0 + \theta_1 + \theta_2 x + \theta_3 m + \theta_4 x + \theta_5 m + \theta_6'c) P(m|R=0,x,c) P(x|R=1,c)$
$= \theta_0 + \theta_1 + (\theta_2 + \theta_4) E[X|R=1,c] + (\theta_3 + \theta_5) E[M|R=0,c] + \theta_6'c$
Similarly,
$E[Y|R=0,c] = \Sigma_{xm} E[Y|R=0,m,x,c] P(m|R=0,x,c) P(x|R=0,c)$
$= \Sigma_{xm} (\theta_0 + \theta_2 x + \theta_3 m + \theta_6'c) P(m|R=0,x,c) P(x|R=0,c)$
$= \theta_0 + \theta_2 E[X|R=0,c] + \theta_3 E[M|R=0,c] + \theta_6'c$
Thus, $\mu_x - E[Y|R=0,c] = \theta_1 + \theta_2 \{E[X|R=1,c] - E[X|R=0,c]\} + \theta_4 E[X|R=1,c] + \theta_5 E[M|R=0,c]$
Also,
$E[Y|R=1,c] = E[Y|R=1,x,m,c] P(x|R=1,c)$
$= \Sigma_{xm} (\theta_0 + \theta_1 + \theta_2 x + \theta_3 m + \theta_4 x + \theta_5 m + \theta_6'c) P(m|R=1,x,c) P(x|R=1,c)$
$= \theta_0 + \theta_1 + (\theta_2 + \theta_4) E[X|R=1,c] + (\theta_3 + \theta_5) E[M|R=1,c] + \theta_6'c$
Thus, $E[Y|R=1,c] - \mu_m = (\theta_3 + \theta_5) \{E[M|R=1,c] - E[M|R=0,c]\}$

Note that

$E[Y|R=1,c] = \Sigma_{xm} E[Y|R=1,m,x,c] P(m|R=1,x,c)P(x|R=1,c)$
$= \Sigma_{xm} (\alpha_0 + \alpha_1 x + \alpha_2 m + \alpha_3'c) P(m|R=1,x,c)P(x|R=1,c)$
$= \alpha_0 + \alpha_1 E[X|R=1,c] + \alpha_2 E[M|R=1,c] + \alpha_3'c$
Similarly we have that:
$E[Y|R=0,c] = \Sigma_x E[Y|R=0,x,c] P(m|R=0,x,c)P(x|R=0,c)$
$= \Sigma_x (\beta_0 + \beta_1 x + \beta_2 m + \beta_3'c) P(m|R=0,x,c)P(x|R=0,c)$
$= \beta_0 + \beta_1 E[X|R=0,c] + \beta_2 E[M|R=0,c] + \beta_3'c$
Thus,
$E[Y|R=1,c] - E[Y|R=0,c]$
$= (\alpha_0 - \beta_0) + \alpha_1 E[X|R=1,c] - \beta_1 E[X|R=0,c] + \alpha_2 E[M|R=1,c] - \beta_2 E[M|R=0,c]$
$= (\alpha_0 - \beta_0) + \alpha_1 E[X|R=1,c] - \beta_1 E[X|R=0,c] + \alpha_2 E[M|R=1,c] - \beta_2 E[M|R=0,c]$
$+ \alpha_1 E[X|R=0,c] - \alpha_1 E[X|R=0,c] + \alpha_2 E[M|R=0,c] - \alpha_2 E[M|R=0,c]$
$= (\alpha_0 - \beta_0) + (\alpha_1 - \beta_1)E[X|R=0,c] + (\alpha_2 - \beta_2)E[M|R=0,c] + \alpha_1\{E[X|R=1,c] - E[X|R=0,c]\} + \alpha_2\{E[M|R=1,c] - E[M|R=0,c]\}$

A so-called detailed Oaxaca-Blinder decomposition would refer to the terms $(\alpha_0 - \beta_0)$ and $(\alpha_1 - \beta_1)E[X|R=0,c]$ and $(\alpha_2 - \beta_2)E[M|R=0,c]$ as the unexplained portion, and then partition the "explained" portion into the part independently explained by X i.e. $\alpha_1\{E[X|R=1,c] - E[X|R=0,c]\}$ and the part independently explained by M i.e. $\alpha_2\{E[M|R=1,c] - E[M|R=0,c]\}$, with all terms summing to equal the total disparity $E[Y|R=1,c] - E[Y|R=0,c]$.

Note that by definition, $(\alpha_0 - \beta_0) = \theta_1$, $\beta_1 = \theta_2$, $\beta_2 = \theta_3$, $(\alpha_1 - \beta_1) = \theta_4$, and $(\alpha_2 - \beta_2) = \theta_5$.

Note also that
$(\alpha_1 - \beta_1)E[X|R=0,c] + \alpha_1\{E[X|R=1,c] - E[X|R=0,c]\}$
$= \alpha_1 E[X|R=1,c] - \beta_1 E[X|R=0,c] + \beta_1 E[X|R=1,c] - \beta_1 E[X|R=1,c]$
$= (\alpha_1 - \beta_1)E[X|R=1,c] + \beta_1\{E[X|R=1,c] - E[X|R=0,c]\}$

Thus, $\mu_m - E[Y|R=0,c]$
$= \theta_1 + \theta_2\{E[X|R=1,c] - E[X|R=0,c]\} + \theta_4 E[X|R=1,c] + \theta_5 E[M|R=0,c]$
$= (\alpha_0 - \beta_0) + \beta_1\{E[X|R=1,c] - E[X|R=0,c]\} + (\alpha_1 - \beta_1)E[X|R=1,c] + (\alpha_2 - \beta_2)E[M|R=0,c]$
$= (\alpha_0 - \beta_0) + \alpha_1\{E[X|R=1,c] - E[X|R=0,c]\} + (\alpha_1 - \beta_1)E[X|R=0,c] + (\alpha_2 - \beta_2)E[M|R=0,c]$
Also,
$E[Y|R=1,c] - \mu_m$
$= (\theta_3 + \theta_5)\{E[M|R=1,c] - E[M|R=0,c]\}$
$= \alpha_2\{E[M|R=1,c] - E[M|R=0,c]\}$

Thus, under assumption A2, the residual disparity under an intervention to equalize M alone is in fact equal to the sum of the "unexplained" portion and the portion "independently explained" by X. The disparity reduction is equal to the portion "independently explained" by M. (Note that these formulae equate to the ones in the main text if $\theta_4 = 0$ such that $\alpha_1 = \beta_1$ and $\theta_5 = 0$ such that $\alpha_2 = \beta_2$).

This result provides some further intuition for why the disparity reduction under Proposition 4 does not generally equal the difference between reductions under Proposition 1 (equalize X alone) and Proposition 3 (equalize X and M). This would only be so under the special case $\alpha_1 = \omega_1$ i.e. M does not mediate the effect of X (and that both assumptions A1 and A2 hold). Only in that special case could the portion "independently explained" by X be interpreted as the disparity reduction under an intervention to equalize X alone (i.e. Proposition 1).